\documentclass[10pt,journal,compsoc]{IEEEtran}

\usepackage{threeparttable}
\usepackage[utf8]{inputenc}
\usepackage{algorithm}
\usepackage{algorithmic}
\usepackage{url}
\usepackage{subfigure}
\usepackage{colortbl}
\usepackage{multirow}
\usepackage{color}
\usepackage{graphicx}
\usepackage{amsmath}
\usepackage{amsfonts}
\usepackage{bm}
\usepackage{makecell}

\ifCLASSOPTIONcompsoc

  \usepackage[nocompress]{cite}
\else
  \usepackage{cite}
\fi

\ifCLASSINFOpdf

\else

\fi

\begin{document}

\title{PBScaler: A Bottleneck-aware Autoscaling Framework for Microservice-based Applications}

\author{Shuaiyu~Xie,
        Jian~Wang,
        Bing~Li,
        Zekun~Zhang,
        Duantengchuan Li,
        Patrick C. K. Hung
\IEEEcompsocitemizethanks{\IEEEcompsocthanksitem S. Xie,  J. Wang, B. Li, Z. Zhang, and D. Li are with the School of Computer Science, Wuhan University, China. P. C. K. Hung is with the Faculty of Business and Information Technology, Ontario Tech University, Canada. J. Wang and B. Li are the corresponding authors. \protect\\
E-mail: \{theory,jianwang,bingli,zekunzhang,dtclee1222\}@whu.edu.cn, patrick.hung@ontariotechu.ca}

\thanks{Manuscript received xxx xx, xxxx; revised xxx xx, xxxx.}}

\markboth{Journal of \LaTeX\ Class Files,~Vol.~14, No.~8, August~2015}%
{Shell \MakeLowercase{\textit{et al.}}: Bare Advanced Demo of IEEEtran.cls for IEEE Computer Society Journals}

\IEEEtitleabstractindextext{%
\begin{abstract}

Autoscaling is critical for ensuring optimal performance and resource utilization in cloud applications with dynamic workloads. 
However, traditional autoscaling technologies are typically no longer applicable in microservice-based applications due to the diverse workload patterns and complex interactions between microservices. 
Specifically, the propagation of performance anomalies through interactions leads to a high number of abnormal microservices, making it difficult to identify the root performance bottlenecks (PBs) and formulate appropriate scaling strategies.
In addition, to balance resource consumption and performance, the existing mainstream approaches based on online optimization algorithms require multiple iterations, leading to oscillation and elevating the likelihood of performance degradation.
To tackle these issues, we propose PBScaler, a bottleneck-aware autoscaling framework designed to prevent performance degradation in a microservice-based application. The key insight of PBScaler is to locate the PBs. Thus, we propose TopoRank, a novel random walk algorithm based on the topological potential to reduce unnecessary scaling. By integrating TopoRank with an offline performance-aware optimization algorithm, PBScaler optimizes replica management without disrupting the online application. Comprehensive experiments demonstrate that PBScaler outperforms existing state-of-the-art approaches in mitigating performance issues while conserving resources efficiently.
\end{abstract}

}

\maketitle

\IEEEdisplaynontitleabstractindextext

\IEEEpeerreviewmaketitle

\ifCLASSOPTIONcompsoc
\IEEEraisesectionheading{\section{Introduction}\label{Introduction}}

\IEEEPARstart{W}{ith} the advancement of microservice architecture, an increasing number of cloud applications are migrating from monolithic architecture to microservice architecture \cite{ma2020automap, kim2013root, shan2019diagnosis, qiu2020firm, liu2018fuzzy, zhang2022robust}. This new architecture reduces application coupling by breaking a monolithic application into multiple microservices that communicate with each other via HTTP or RPC protocols \cite{lin2018microscope}. Moreover, each microservice can be developed, deployed, and scaled independently by separate teams, enabling rapid application development and iteration. Nevertheless, the unpredictability of external workloads and the complexity of interactions between microservices can result in performance degradation \cite{liu2021microhecl, choi2021phpa, kwan2019hyscale}. Cloud providers must prepare excessive resources to meet the service level objective (SLO) of application owners, which usually causes unnecessary waste of resources \cite{baarzi2021showar, park2021graf}. As a result, the imbalance between satisfying SLO and minimizing resource consumption becomes a major challenge encountered by resource management in microservices.

Microservice autoscaling refers to the capability of allocating resources elastically in response to workload variations \cite{al2017elasticity}. By utilizing the elasticity property of microservices, autoscaling can mitigate the conflict between resource cost and performance. However, the autoscaling of microservices suffers from accurately scaling the performance bottleneck (PB) in a short period. 
Due to the complexity of communication between microservices, the performance degradation of a PB may propagate to other microservices via message passing \cite{kim2013root}, resulting in a high number of abnormal microservices at the same time. We demonstrated this by injecting burst workloads to specific microservices in Online Boutique \footnote{https://github.com/GoogleCloudPlatform/microservices-demo}, an open-source microservice application developed by Google. Fig. \ref{fig:part-dependency} shows that the performance degradation in the PB \textit{Recommend} can spread to the upstream microservices like \textit{Checkout} and \textit{Frontend}. To further verify the importance of accurately scaling the PB, we conducted stress testing and scaled different microservices separately. As shown in Fig. \ref{fig:root-test}, abnormal microservice (\textit{Frontend}) scaling cannot alleviate SLO violations. However, when we identified and scaled the PB \textit{Recommend}, the performance of the microservice application improved. Unfortunately, locating PBs is usually time-consuming and can occasionally make mistakes  \cite{chen2016causeinfer}.

In recent years, several approaches have been proposed to identify critical microservices before autoscaling. For example, the default autoscaler of Kubernetes \footnote{https://github.com/kubernetes/autoscaler} filters microservices for direct scaling based on a static threshold of computing resources. Yu \textit{et al.} \cite{yu2019microscaler} defined the boundaries of elastic scaling by calculating the service power, which is the ratio between the 50th percentile response time (P50) and the 90th percentile response time (P90). Furthermore, Qiu \textit{et al.} \cite{qiu2020firm} introduced an SVM-based approach for extracting critical paths by analyzing the ratio of various tail latencies. Although these studies have narrowed the scope of autoscaling, they still take into account non-bottleneck microservices that may affect scaling strategies, especially when a large number of microservices in the application are abnormal at the same time. Consequently, there is an urgent need to pinpoint bottleneck microservices accurately before autoscaling.

To balance resource consumption and performance, existing works have employed online optimization algorithms to find near-optimal autoscaling strategies. However, due to the vast range of possible strategies for autoscaling, these approaches require a significant number of attempts, which will be problematic for online applications. For example, Train Ticket\footnote{https://github.com/FudanSELab/train-ticket} is the largest open-source microservice application, consisting of nearly 40 microservices. Assuming that each microservice can have up to 15 replicas, determining the optimal allocation strategy for this application is undoubtedly an NP-hard problem, as there are a maximum of $15^{40}$ scaling alternatives. Additionally, the duration of the feedback loop in online optimization is too long to achieve model convergence. It is also essential to consider the potential risks of performance degradation caused by online optimization. Fig. \ref{fig:microscaler-wave} illustrates the impact of burst workloads on the replica fluctuation and latency fluctuation of  MicroScaler \cite{yu2019microscaler}, an online autoscaling approach incorporating online Bayesian optimization to find the global minimizer of the total cost. The frequent online attempts to create replicas (Fig. \ref{fig:microscaler-wave}a) caused by online optimization result in oscillations and performance degradation (Fig. \ref{fig:microscaler-wave}b). As a result, we are inspired to design an offline optimization process fueled by feedback from a simulator.

\begin{figure}[t]
\centering{
 \includegraphics[width=\linewidth]{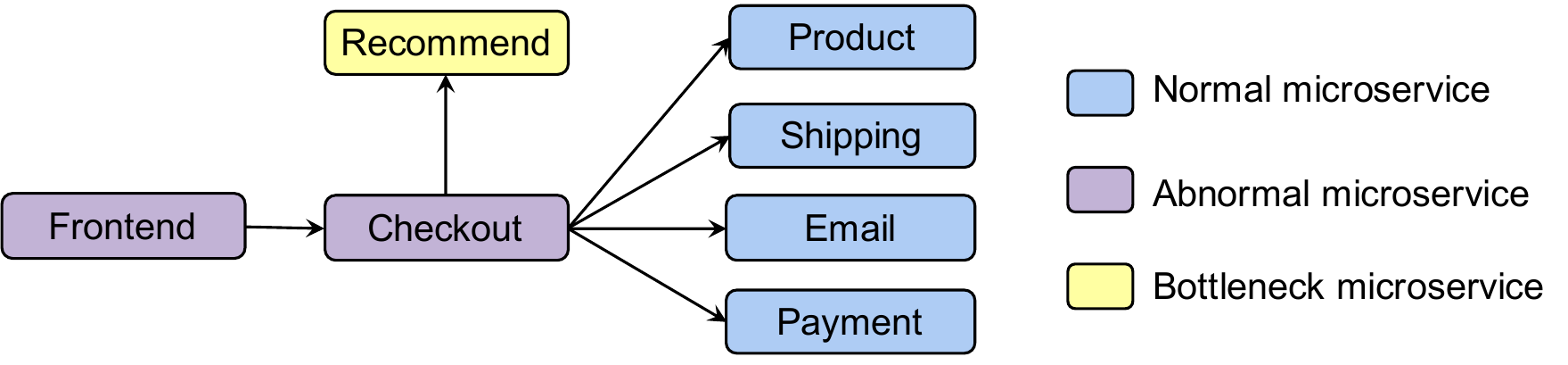}
}
\caption{Part of the invocation relationship in Online Boutique.}
\label{fig:part-dependency}
\end{figure}

\begin{figure}[t]
\centering{
 \includegraphics[width=0.9\linewidth]{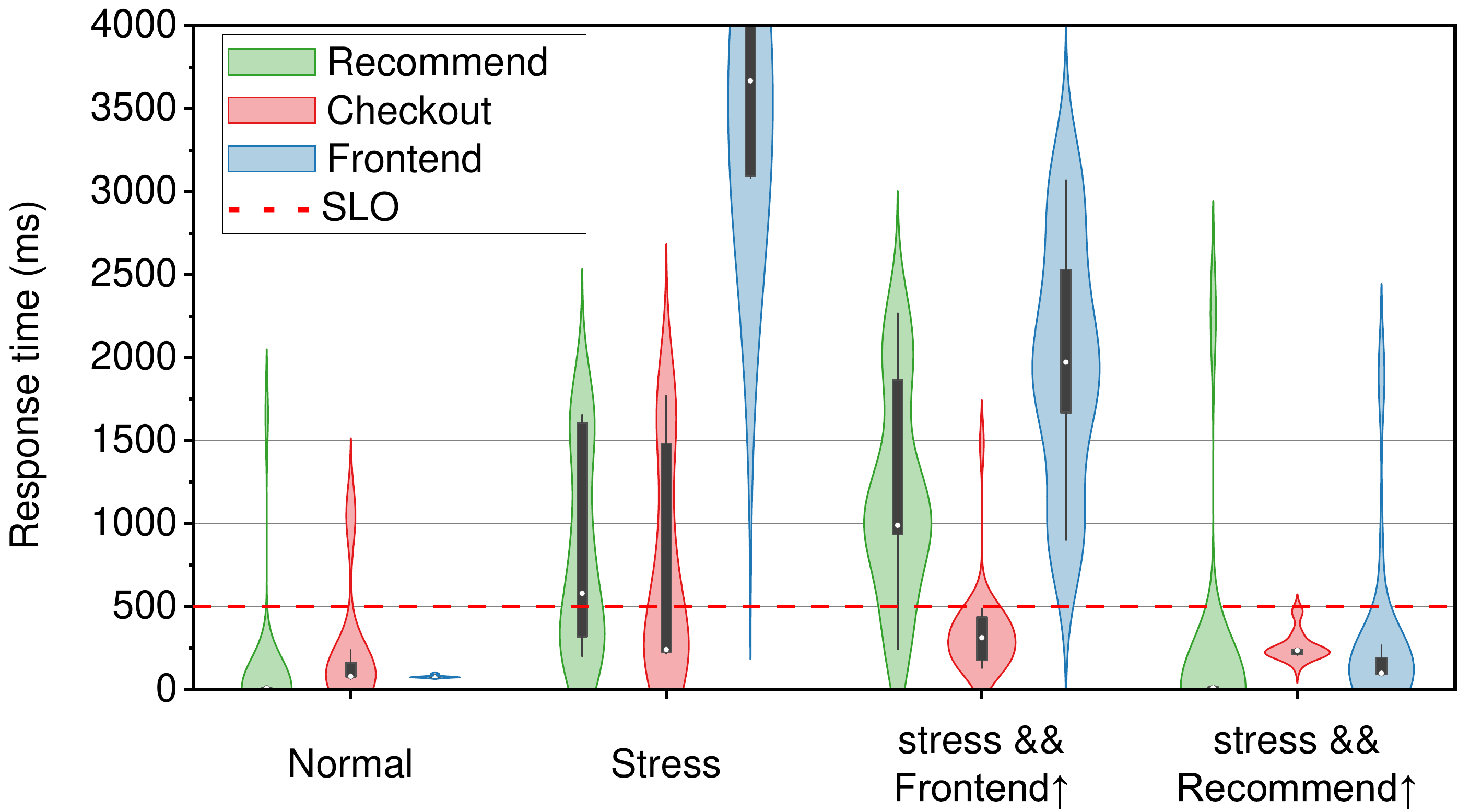}
}
\caption{Latency distribution in four scenarios.}
\label{fig:root-test}
\end{figure}


This paper presents PBScaler, a horizontal autoscaling framework designed to prevent performance degradation in microservice-based applications by identifying and addressing bottlenecks. Instead of optimizing resources for all abnormal microservices, as was done in previous work \cite{yu2019microscaler, baarzi2021showar}, we propose TopoRank, a random walk algorithm based on topological potential theory (TPT) to identify performance bottlenecks (PBs). By taking into account microservice dependencies and anomaly potential, TopoRank improves the accuracy and explainability of bottleneck localization. 
After identifying the PBs by TopoRank, PBScaler further employs a genetic algorithm to find nearly optimal strategies. To avoid application oscillation caused by excessive optimization, the process is conducted offline and is guided by an SLO violation predictor, which simulates the online application and provides feedback to the scaling strategies. The main contributions of the paper are summarized as follows:

\begin{itemize}
\item We propose PBScaler, a bottleneck-aware autoscaling framework designed to prevent performance degradation in a microservice-based application.
By pinpointing bottlenecks, PBScaler can reduce unnecessary scaling and expedite the optimization process.

\item 
We employ a genetic algorithm-based offline optimization process to optimize resource consumption while avoiding SLO violations. This process is guided by an SLO violation predictor and is designed to strike a balance between resource consumption and performance without disrupting online applications.

\item We design and implement PBScaler in the Kubernetes system. To evaluate its effectiveness, we conduct extensive experiments with real-world and emulated workload injection on two widely-used microservice systems running in an online environment. Experimental results demonstrate that PBScaler outperforms several state-of-the-art elastic scaling methods.

\end{itemize}

The rest of the paper is organized as follows. In Section 2, we discuss the related work about bottleneck analysis and autoscaling for microservices. In Section 3, we describe the overall system in detail. In Section 4, we present the evaluations and experimental results. Section 5 concludes our work and discusses the future research direction.

\begin{figure}[t]
\centering{
 \subfigure[ Replica fluctuation ]{
 \includegraphics[width=0.47\linewidth]{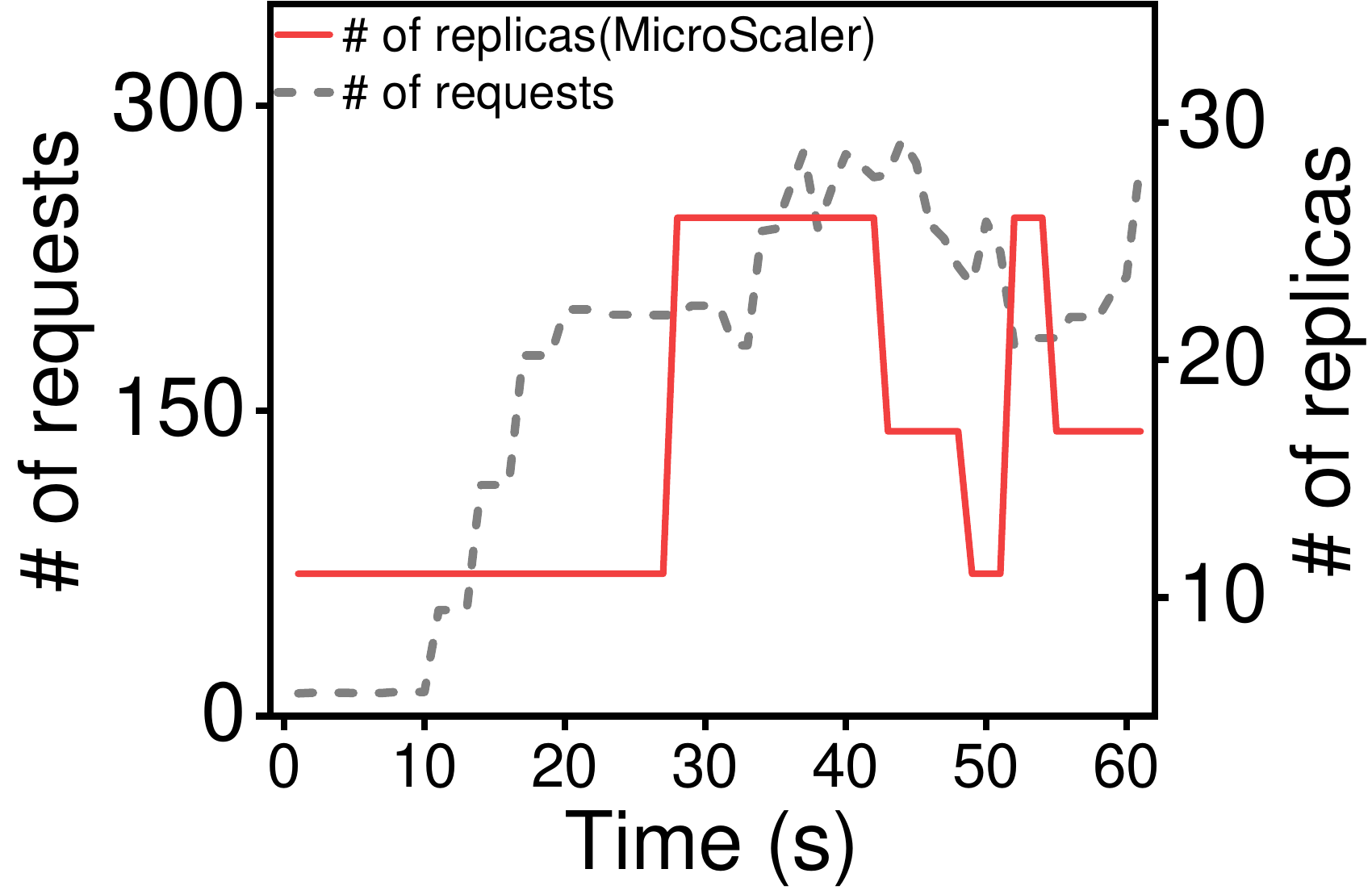} } 
 \subfigure[ Latency fluctuation ]{  \includegraphics[width=0.48\linewidth]{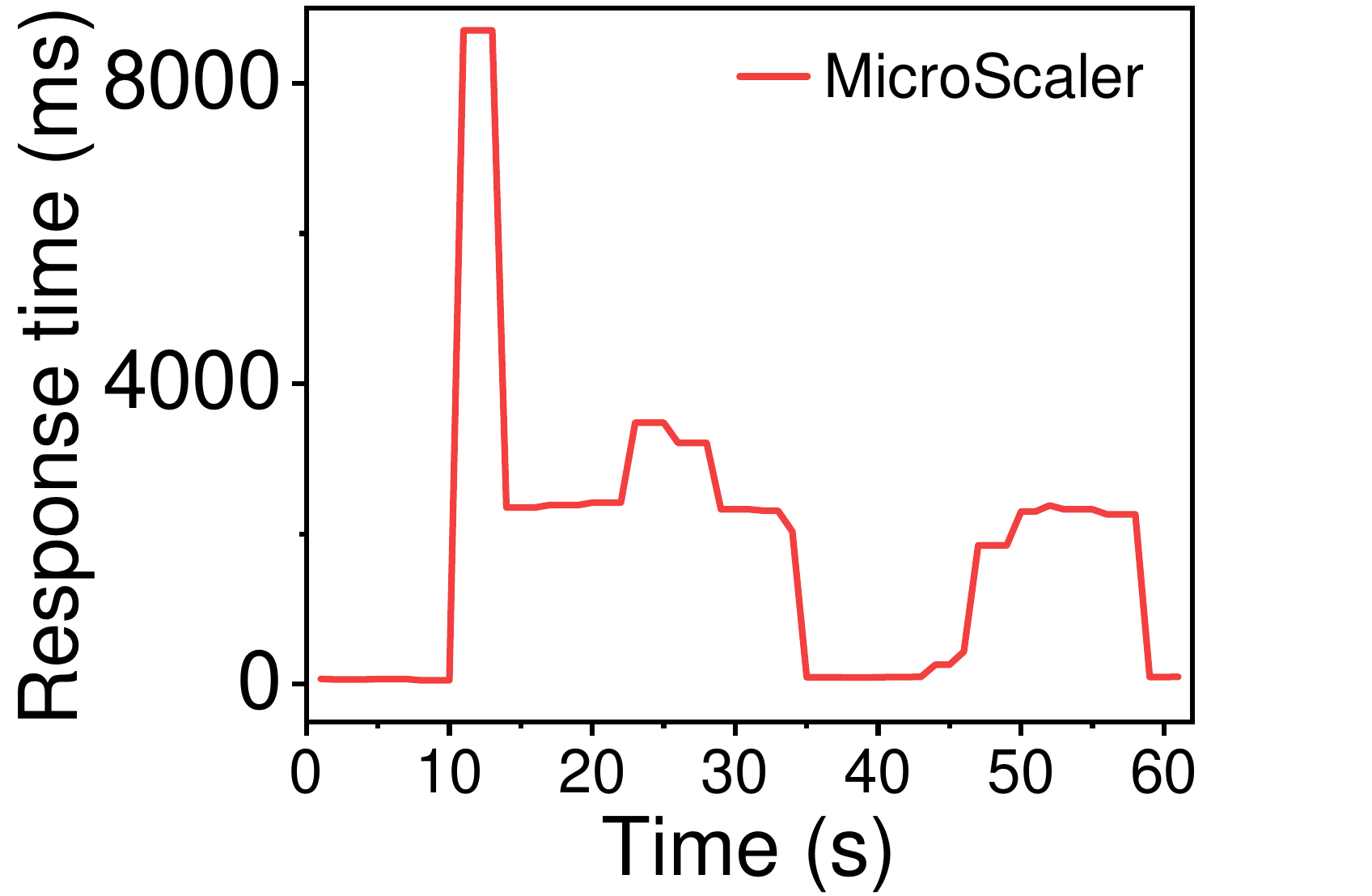}}
}
\caption{The replica fluctuation and latency fluctuation of MicroScaler under burst workloads. Excessive online attempts (a) cause oscillations and performance degradation (b).}
\label{fig:microscaler-wave}
\end{figure}

\section{Related Work}

With the advancement of cloud computing, numerous autoscaling methods for cloud resources, such as virtual machines or containers, have been proposed in academia and industry \cite{sharma2011cost, da2015autoelastic, roy2011efficient, wang2022trust}. However, 
autoscaling for microservices can be much more complicated due to the intricate dependencies between microservices.

Performance bottleneck analysis, also known as root cause analysis, is a useful way for quickly locating the bottleneck responsible for the performance degradation of microservices, hence reducing the time and effort required for autoscaling. In this section, we will analyze the related work on bottleneck analysis and autoscaling for microservices.

\begin{figure*}
	\centering
	\includegraphics[width=0.95\textwidth]{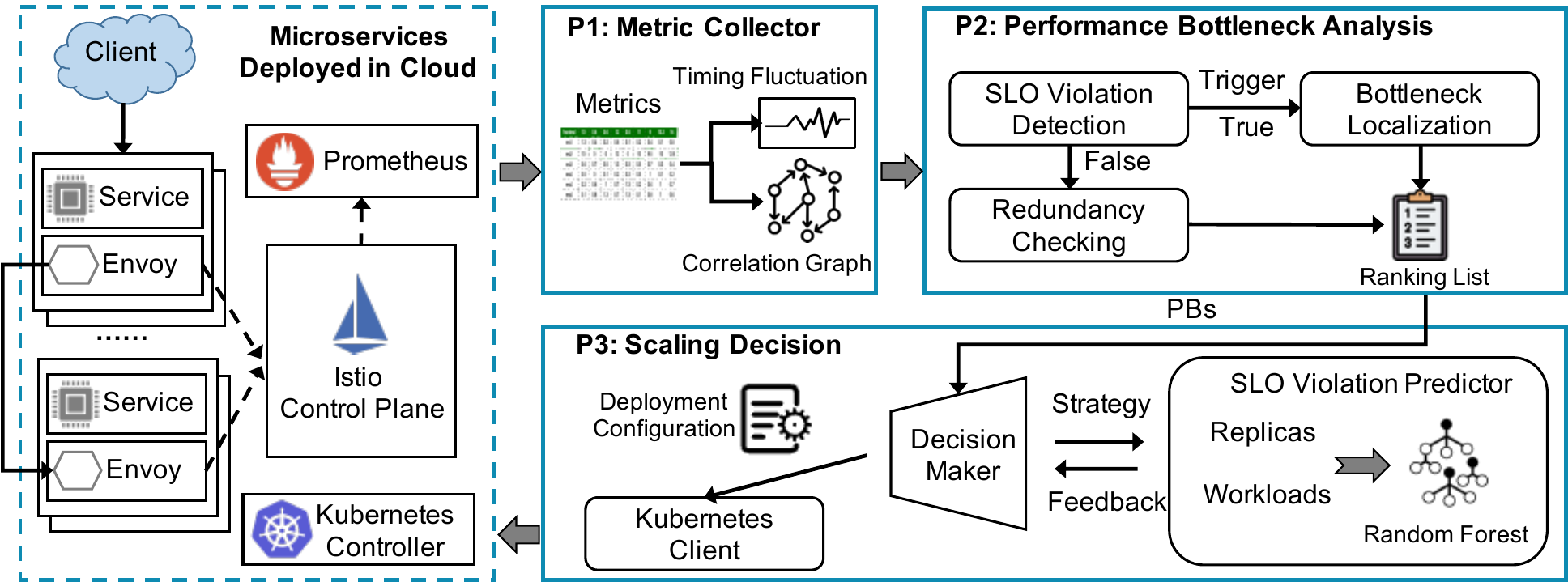}
	\caption{Framework of PBScaler.}
	\label{fig:structure}
 \vspace{-2mm}
\end{figure*}

\subsection{Bottleneck Analysis}

In recent years, numerous methods for bottleneck analysis in microservice scenarios have been developed, most of which rely on three types of data: logs, traces, and metrics. 1) Logs. Jia \textit{et al.} \cite{jia2017approach} and Nandi \textit{et al.} \cite{nandi2016anomaly} first extracted templates and flows from normal-state logs, matched them with target logs, and filtered out abnormal logs. 2) Traces. Trace is an event tracking-based record that reproduces the request process between microservices. Several studies \cite{yu2021microrank, yu2021tracerank, mi2013toward, li2021practical} have been introduced to pinpoint the bottlenecks using traces. Yu \textit{et al.} \cite{yu2021microrank, yu2021tracerank} located bottlenecks by combining spectrum analysis and the PageRank algorithm on the dependency graph constructed by traces, while Mi \textit{et al.} \cite{mi2013toward} presented an unsupervised machine learning prototype to learn the pattern of microservices and filter out abnormal microservices. However, using traces can be intrusive to the code and requires operators to have a deep understanding of the structure of the microservices. 3) Metrics. Some approaches \cite{zhangaamr, wu2020microrca, kim2013root} leverage graph random walk algorithms to simulate the propagation process of anomalies and then find bottlenecks by integrating statistical features of metrics and dependencies between microservices. Additionally, methods such as CauseInfer \cite{chen2016causeinfer} and MicroCause \cite{meng2020localizing} focused on building metrics causality graphs with causal inference, which typically involve hidden indirect relationships between metrics. 

Since the workflow code is rarely modified when monitoring metrics, collecting metrics for microservices is usually cheaper than using trace. Moreover, using metrics as the primary monitoring data can reduce the cost of integrating bottleneck analysis and autoscaling, as metrics are widely used in the latter scenario. Despite these approaches' advantages, most have no preference in selecting the starting point for abnormal backtracking. In contrast, our approach begins random walks from microservices with greater anomaly potential, accelerating convergence speed and improving bottleneck localization accuracy.

\subsection{Autoscaling for Microservices}

Existing autoscaling methods for microservices can be categorized into five groups. 1) Rule-based heuristic approach. KHPA, Libra \cite{balla2020adaptive}, KHPA-A \cite{casalicchio2017auto}, and PEMA \cite{hossen2022lightweight} manage the number of microservice replicas based on resource thresholds and specific rules. However, since different microservices have varying sensitivities to specific resources,  expert knowledge is needed to support autoscaling for these various microservices. 2) Model-based approach. Microservices can be modeled to predict their status under particular configurations and workloads. Queuing theory \cite{baresi2020simulation, tong2021holistic} and graph neural network (GNN) \cite{park2021graf} are commonly used to build performance prediction models for microservices. 3) Control theory-based approach \cite{baarzi2021showar, baresi2020simulation}. Using the control theory, SHOWAR \cite{baarzi2021showar} dynamically adjusts the microservice replicas to correct the error between monitoring metrics and thresholds. 4) Optimization-based approach. These methods \cite{yu2019microscaler, gias2019atom} make a large number of attempts to find the optimal strategy given the present resources and workloads. The key to these approaches is to reduce the decision-making scope to speed up the process. 5) RL-based Approach. Reinforcement learning (RL) has been widely used in resource management for microservices. MIRAS \cite{yang2019miras} adopts a model-based RL method for decision-making to avoid the high sampling complexity of the real environment. FIRM \cite{qiu2020firm} leverages a support vector machine (SVM) to identify the critical path in microservices and a deep deterministic policy gradient (DDPG) algorithm to make hybrid scaling strategies for microservices along the path. RL-based methods require constant interactions with the environment during exploration and are incapable of adapting to the dynamic microservices architecture. 

In conclusion, while the aforementioned autoscaling techniques have their respective advantages, they pay little attention to performance bottlenecks. Consuming computer resources for non-bottleneck microservices will inevitably increase scaling costs and lengthen decision-making. Our method, on the other hand, focuses on locating performance bottlenecks.

\section{System Design}
 
We present PBScaler, a PB-centric autoscaling controller, to locate PBs and optimize replicas for them. As shown in Fig.~\ref{fig:structure}, PBScaler comprises three components: 1) \textit{Metric Collector}: To provide real-time insights into the applications' status, we design a metric collector that captures and integrates monitoring metrics from Prometheus\footnote{ https://prometheus.io} at fixed intervals. 2) \textit{Performance Bottleneck Analysis}: With the assistance of the \textit{metric collector}, this component performs SLO violation detection and redundancy checking to identify microservices with abnormal behavior. Next, the bottleneck localization process will be triggered to pinpoint the PBs in the abnormal microservices. 3) \textit{Scaling Decision}: This component aims to determine the optimal number of replicas for  PBs using an evolutionary algorithm. Finally, PBScaler generates configuration files with optimized strategies and commits them to the kubernetes-client\footnote{https://github.com/kubernetes-client/python}, which regulates the replica count of microservices.

\begin{table}[t]
	\caption{Labels of metrics collected from the monitoring tools. 
}
 \label{tab:metric-labels}
	\begin{tabular}{ll}
		\Xhline{0.8pt}
        \textbf{Level} & \textbf{Label} \\
        \hline
		 &kube\_pod\_info\\
		&container\_cpu\_usage\_seconds\_total\\
		&container\_memory\_usage\_bytes\\ 
		&container\_spec\_cpu\_quota\\
		&container\_spec\_memory\_limit\_bytes\\ 
		Container Level&container\_fs\_usage\_bytes\\ 
		&container\_fs\_write\_seconds\_total\\
		&container\_fs\_read\_seconds\_total\\ &container\_network\_receive\_bytes\_total\\
		&container\_network\_transmit\_bytes\_total\\
		\hline
		&istio\_request\_duration\_milliseconds\_bucket\\
		Microservice Level&istio\_requests\_total\\
		&istio\_tcp\_received\_bytes\_total\\
		\Xhline{0.8pt}
	\end{tabular}
\end{table}

\subsection{Metric Collector}

The autoscaling controller relies on observability for microservice applications, such as system load, tail latency, and invocation relationships between microservices, to determine whether elastic scaling should be performed and how many resources should be allocated. While a trace-based monitor can reflect the dependencies and performance of microservices, it requires a deep understanding of the program and code injection \cite{lin2018microscope}. Moreover, real-time analysis of massive tree-structural traces demands a considerable amount of processing time. Therefore, we design a metric-based monitor, the \textit{Metric Collector}, based on non-intrusive service mesh technology to minimize disruptions to business flows. As shown in Table \ref{tab:metric-labels}, PBScaler uses Prometheus and kube-state-metrics to gather and categorize these metrics ($M$), including tail latency, invocation relationships between microservices, resource consumption, and microservice workload. For example, \textit{container\_cpu\_usage\_seconds\_total} is a resource metric that records the Central Processing Unit (CPU) usage at the container level. \textit{istio\_requests\_total} records the TCP requests between microservices. Fig. \ref{fig:grafana} displays a visualization of the query results for \textit{istio\_requests\_total}, where each point represents the average request rate within the preceding minute for a specific invocation relationship (e.g., \textit{frontend}$\to$\textit{currency} as marked in Fig. \ref{fig:grafana}). This information serves as a means to intuitively reflect the workload on microservices. Furthermore, \textit{istio\_requests\_total} also records all invocation relationships (listed in the legend of Fig. \ref{fig:grafana}), which can be utilized to construct a microservice correlation graph $\mathcal{G}_{c}$ similar to the one depicted in Fig. \ref{fig:abnomal_propagate}. Meanwhile, \textit{istio\_request\_duration\_milliseconds\_bucket} serves as a performance metric reflecting the tail latency of a microservice. Typically, we collect the P90 tail latency of each microservice to observe their performance. The monitoring interval of Prometheus is set to five seconds, and the collected metrics data is stored in a time-series database.

\textbf{Service Mesh}. A service mesh is an infrastructure that enables developers to add advanced features, such as observability and traffic management, to cloud applications without requiring additional code. One popular open-source service mesh implementation is Istio\footnote{https://istio.io}, designed to seamlessly integrate with Kubernetes. When a pod starts up in Kubernetes, Istio launches an envoy proxy within the pod to intercept network traffic, enabling workload balancing and monitoring.

\begin{figure}
    \centering
    \includegraphics[width=\linewidth]{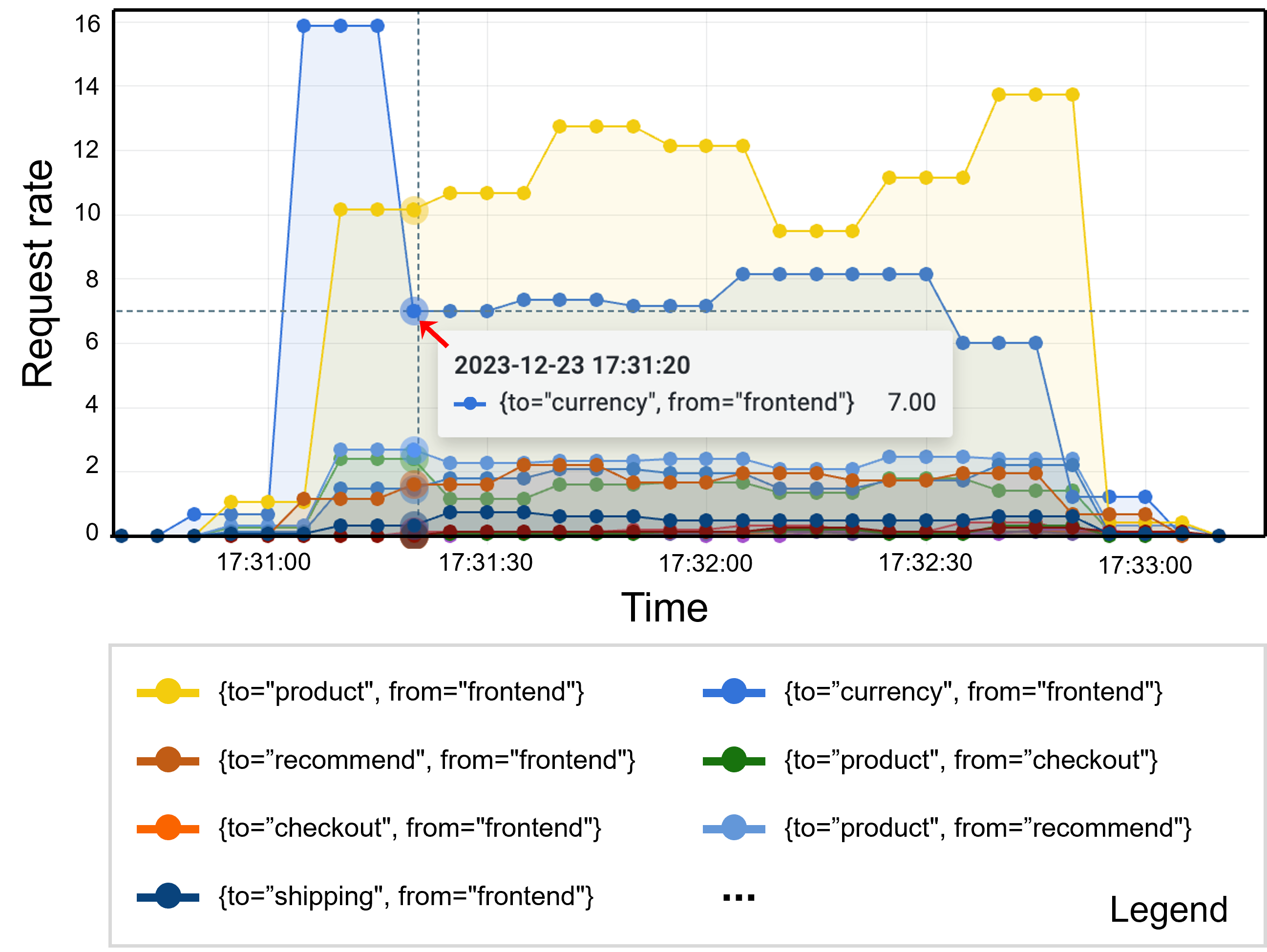}
    \caption{Visualization of query results for metric \textit{istio\_requests\_total}.}
    \label{fig:grafana}
\end{figure}
\vspace{-2mm}

\subsection{Performance Bottleneck Analysis}

\textit{Performance Bottleneck Analysis} (PBA) is a process designed to discover performance degradation and resource waste in microservice applications to infer PBs of the current problem. As stated in Section \ref{Introduction}, by accurately locating these bottlenecks, PBA can enhance the performance of autoscaling and reduce excessive resource consumption. The PBA process in PBScaler is depicted in Algorithm \ref{al:PBA}.

\subsubsection{SLO Violation Detection} 
To detect abnormalities in microservices, PBScaler uses service level objectives (SLOs) to compare with specific metrics. It is considered abnormal if a microservice has numerous SLO violations, i.e., performance degradation. As discussed in \cite{wu2020microrca, chen2016causeinfer}, detecting SLO violations is a critical step in triggering bottleneck localization. The invocation relationships collected by the \textit{Metric Collector} can be leveraged to build a microservice correlation graph $\mathcal{G}_{c}$. PBScaler inspects the P90 tail latency of all invocation edges in $\mathcal{G}_{c}$ every 15 seconds to timely detect performance degradation. If the tail latency of an invocation exceeds a predetermined threshold (such as the SLO), the invoked microservice of the invocation will be added to the set of abnormal microservices ($\mathbb{S}$), and the bottleneck localization process will be activated. To account for occasional noise in the microservice latency, the threshold is set to SLO$ \times (1 + \frac{\alpha}{2})$, where $\alpha$ is used to adjust the tolerance to noise.

\begin{algorithm}[t]
	\renewcommand{\algorithmicrequire}{\textbf{Input:}}
	\renewcommand{\algorithmicensure}{\textbf{Output:}}
	\caption{Performance Bottleneck Analysis}
	\label{al:PBA}
	\begin{algorithmic}[1]
		\REQUIRE SLO \\
 microservice correlation graph $\mathcal{G}_{c}$ \\
  confidence level $cl$ \\ 
  tolerance to noise $\alpha$ \\
  degree of significance $\beta$ \\
  impact factor $\sigma$
		\ENSURE ranking list $\bm{rl}$
        \STATE Initialize $\bm{rl}$ and abnormal microservice set $\mathbb{S}$
        
		\FOR{each $v_{i}$ in $\mathcal{G}_{c}$.nodes}
		\FOR{each $e_{j,i}$ in in-edges of $v_{i}$}
        \STATE /* SLO violation detection */
		\IF {$P90(e_{j,i}) \textgreater $ SLO $ * (1 + \alpha / 2)$}
		\STATE Store $v_{i}$ in $\mathbb{S}$
		\ENDIF
		\ENDFOR
		\ENDFOR
		\IF {$\mathbb{S}$ is empty}
        \STATE /* Redundancy checking */
        		\FOR{each $v_{i}$ in $\mathcal{G}_{c}.nodes$}
          \STATE $\bm{w_p}^i$, $\bm{w_c}^i$ $\gets$ Get workloads for microservice $i$ from the Metric Collector
        		\STATE $t$, $P$ $\gets$ $ttest$($\bm{w_c}^i$, $\bm{w_p}^i * \beta$)
        		\IF {$t< 0$ and $P < cl$}
        		\STATE Store $v_{i}$ in $\bm{rl}$
        		\ENDIF
          	\STATE Return $\bm{rl}$	\ENDFOR
		\ENDIF
  \STATE $\bm{rl}$ = $TopoRank$($\mathbb{S}$)
  \STATE Return $\bm{rl}$
	\end{algorithmic}  
\end{algorithm}

\subsubsection{Redundancy Checking}

In the absence of performance anomalies, some microservices may be allocated more resources than required. However, identifying such cases can be difficult through metrics alone, potentially leading to wasting limited hardware resources. To avoid this, it is essential to identify which microservices have allocated excess resources. PBScaler uses the rate of workload change per second of microservices to determine whether resources are redundant. This strategy is more effective than relying only on resource consumption because different microservices may have varying sensitivity to heterogeneous resources. The main idea behind redundancy checking is to employ hypothesis testing to detect whether a microservice's current workload $\bm{w_c}^i$ is significantly lower than its past workload (denoted as $\bm{w_p}^i$). The degree of significance is adjusted by the parameter $\beta$, and the hypothesis test can be formulated as:

\begin{equation}\label{eq:hypothesis-test}
	\left \{
	\begin{aligned}
		H_{0},  & \quad \bm{w_c}^i \geq \bm{w_p}^i \times \beta \\
		H_{1},  & \quad   \bm{w_c}^i < \bm{w_p}^i \times \beta.
	\end{aligned}
	\right.
\end{equation}

To perform the hypothesis test, we first fetch the current and historical workloads of the target microservices from the \textit{Metric Collector}. We then use a one-sided test to compute the p-value $P$. If $P$ does not exceed the confidence level $cl$ (which is set to 0.05 by default), we reject the null hypothesis $H_{0}$ and consider the microservice $i$ to have redundant resources. 

\begin{algorithm}[t]
\renewcommand{\algorithmicrequire}{\textbf{Input:}}
\renewcommand{\algorithmicensure}{\textbf{Output:}}
	\caption{TopoRank}
	\label{al:TopoRank}
	\begin{algorithmic}[1]
		\REQUIRE abnormal subgraph $\mathcal{G}_{a}$ \\ 
  impact factor $\sigma$ \\
        preference vector $\bm{u}$ \\
        transition matrix $\textbf{\textit{P}}$ \\
        collected metrics $M$ \\
		\ENSURE ranking list $\bm{rl}$
  \STATE Initialize $\bm{rl}$ and preference vector $\bm{u}$
  
  \FOR{each $v_{i}$ in $\mathcal{G}_{a}$.nodes}
  \STATE $a_i \gets$ Anomaly degree of $v_i$
  \STATE $\varphi \gets a_i$
            \FOR{each $v_{j}$ in upstream microservices of $v_{i}$}
            \STATE $h_{ji} \gets$ Minimum number of hops from $v_{j}$ to $v_{i}$
            \STATE $a_j \gets$ Anomaly degree of $v_j$
            \STATE $\varphi \gets \varphi + a_je^{-(h_{ji} / \sigma)^2}$
            
            \ENDFOR
\STATE $\bm{u}_{i}\gets \varphi$ 
            \STATE $L_t$ $\gets$
    		An array of tail latency for $v_{i}$
		\FOR{each $v_{j}$ in out-edges of $v_{i}$}
        \STATE $L_m \gets$ An array of metric $m$ for $v_{j}$
		\STATE $ \textbf{\textit{P}}_{i, j} \gets \mathop{\max}\limits_{m\in M}(corr(L_t, L_m))$
		\ENDFOR
            
        \ENDFOR
    		
		\STATE $\bm{rl}$ $\gets$ $pageRank$($\textbf{\textit{P}}$, $\bm{u}$)
  \STATE Return $\bm{rl}$
	\end{algorithmic}  
\end{algorithm}


\begin{figure}[t]
  \centering
  \includegraphics[width=\linewidth]{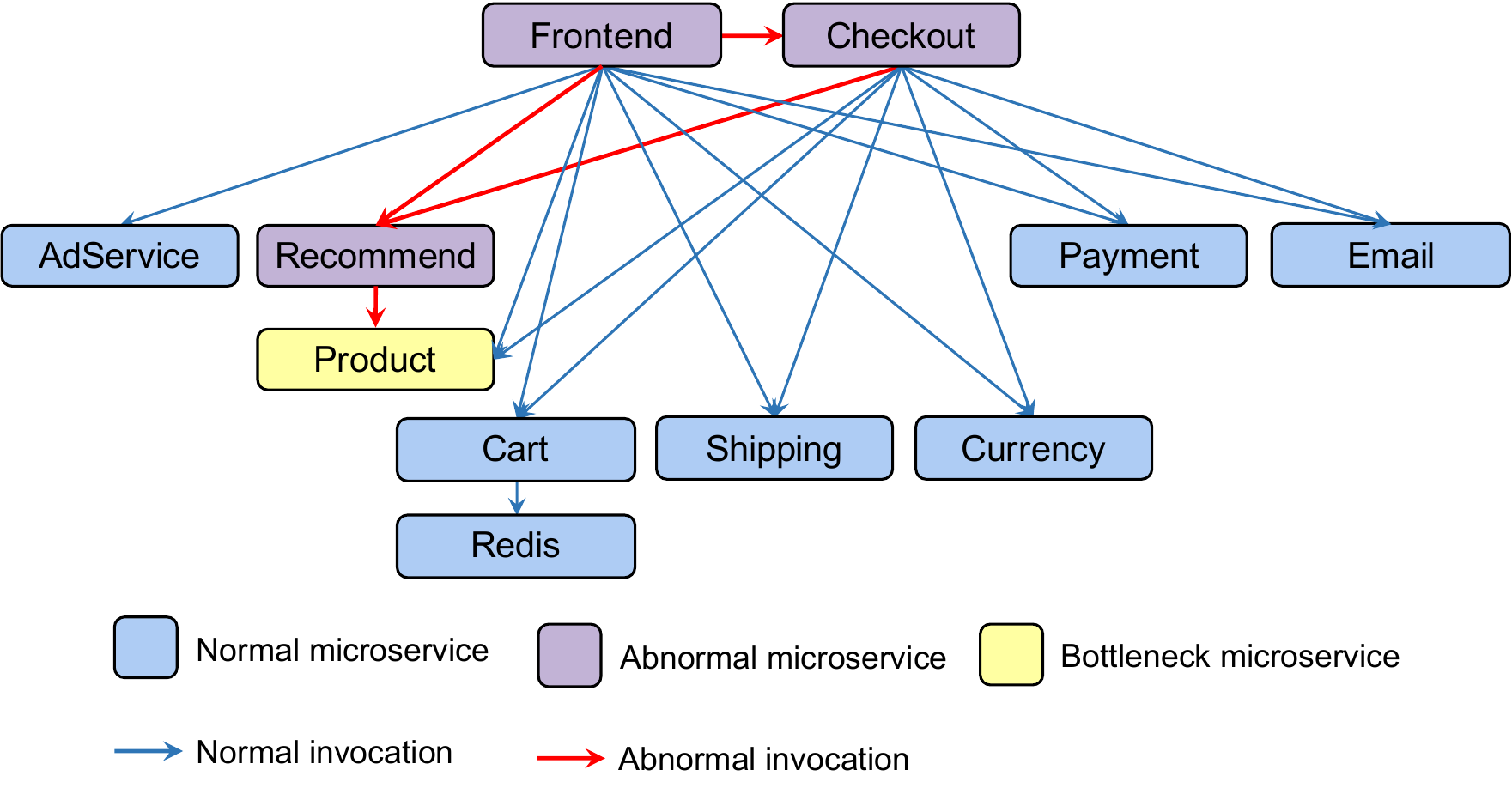}
  \caption{Example of anomaly propagation in microservices.}
\label{fig:abnomal_propagate}
\end{figure}

\begin{figure*}
    \centering
    \includegraphics[width=0.9\linewidth]{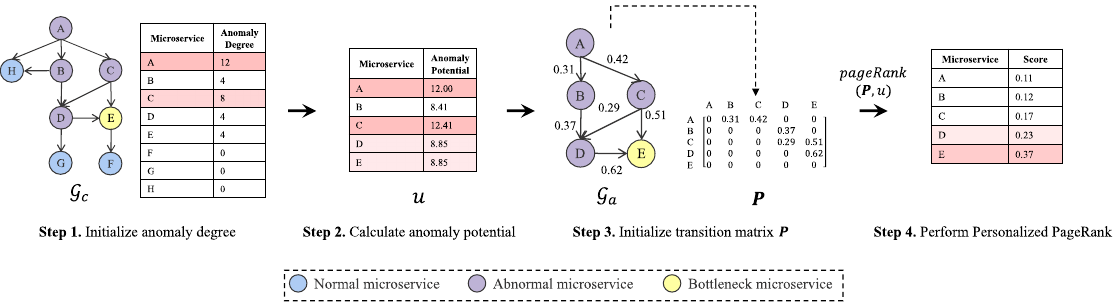}
    \caption{Illustration of the computation process for the TopoRank algorithm.}
    \label{fig:toporank}
\end{figure*}

\subsubsection{Bottleneck Localization}
Because of the complex interactions in a microservice application \cite{zhou2019latent}, not every abnormal microservice needs to be scaled up. For example, Fig. \ref{fig:abnomal_propagate} illustrates how the performance degradation of a bottleneck microservice (e.g., \textit{Product}) can propagate to its upstream microservices (e.g., \textit{Recommend}, \textit{Frontend}, and \textit{Checkout}) along the invocation chains, even if the upstream microservices are not overloaded. Therefore, only the bottleneck microservice must be scaled up while the other abnormal microservices are merely implicated. To pinpoint the bottleneck microservice, we introduce the concept of anomaly potential, which aggregates the anomaly impact of all microservices in a given position. Generally, a performance bottleneck tends to exhibit a high anomaly potential in a microservice application, as it is frequently surrounded by abnormal neighbors that are influenced by the bottleneck. Based on the anomaly potential, we model the backtracking of anomalies in microservice applications as a random walk on a directed graph. Therefore, We design a novel bottleneck localization algorithm, TopoRank, which introduces the topological potential theory (TPT) in random walks to calculate the scores for all abnormal microservices and finally outputs a ranking list ($\bm{rl}$).
The microservices with the highest scores in the $\bm{rl}$ can be recognized as the PBs.

TPT, which originated from the concept of "field" in physics, 
 has been widely used in various works \cite{hu2010topological,mao2020tps} to measure the mutual influence between nodes in complex networks. Since the microservice correlation graph can also be viewed as a complex network, we use TPT to evaluate the anomaly potential of microservices. Specifically, we have observed that in a microservice correlation graph $\mathcal{G}_{c}$, the microservices closer to the PBs, i.e., those with fewer hops, are more likely to be abnormal, as they often have frequent direct or indirect invocations with the PBs. Based on this observation, we evaluate the anomaly potential of microservices using the TPT. To do this, we first extract the abnormal subgraph $\mathcal{G}_{a}$ by identifying the abnormal microservices and their invocation relationships in $\mathcal{G}_{c}$. We then calculate the anomaly potential $\varphi$ for microservice $v_i$ in the abnormal subgraph $\mathcal{G}_{a}$ using the TPT:
 \vspace{-2mm}
\begin{equation}
\label{eq:topo}
    \varphi = a_i + \sum_{j=1}^N{a_j e^{-(h_{ji} / \sigma)^2}},
\end{equation}
where $N$ is the number of upstream microservices of $v_i$ and $a_j$ represents the anomaly degree of $v_j$. PBScaler defines the anomaly degree as the number of SLO violations for a microservice in a time window. $h_{ji}$ denotes the minimum number of hops required from $v_j$ to $v_i$. We use the impact factor $\sigma$ (1 by default) to control the influence range of a microservice. 

Fig. \ref{fig:toporank} illustrates the computation process of the TopoRank algorithm. PBScaler checks the number of SLO violations for each microservice in Step 1 to initialize the anomaly degree. It can be observed that the microservices A and C, positioned upstream of the performance bottleneck (i.e., microservice E), have the highest anomaly degrees, with values of 12 and 8, respectively. This is because anomalies in downstream microservices propagate and accumulate in upstream microservices. Consequently, it is inadvisable to locate PBs solely based on the anomaly degree. In Step 2, PBScaler calculates the anomaly potential for each microservice according to Eq. \ref{eq:topo}. Taking the microservice C as an example, its anomaly potential $\varphi$ can be calculated as: $8+12\times e^{-(1/1)^2}=12.41$.




However, microservices with high anomaly potential values are not necessarily PBs, since anomalies are usually propagated along the microservice correlation graph.  In Step 2 of Fig. \ref{fig:toporank}, the upstream microservice A exhibits the highest anomaly potential due to the propagation of downstream anomalies. Hence, relying solely on the TPT is insufficient for diagnosing PBs. To address this issue, PBScaler incorporates the Personalized PageRank algorithm \cite{jeh2003scaling} to reverse the anomaly propagation on the abnormal subgraph $\mathcal{G}_{a}$ and locate PBs.
Let $\textbf{\textit{P}}$ be the transition matrix of $\mathcal{G}_{a}$ and $\textbf{\textit{P}}_{i, j}$ be the probability of anomaly tracking from $v_i$ to its downstream node $v_j$. Given $v_i$ with out-degree $d$, the standard Personalized PageRank algorithm sets $\textbf{\textit{P}}_{i, j}$ as:
\begin{equation}
\textbf{\textit{P}}_{i, j} = \frac{1}{d}, 
\end{equation}
which means that the algorithm is not biased toward any downstream microservice. However, this definition fails to consider the association between the downstream microservices and the anomaly of the current microservice. Consequently, PBScaler adapts the calculation by giving more attention to the downstream microservices whose metrics are more relevant to upstream response time. For each microservice $v_i$, PBScaler logs a tail latency array ($L_t$). For any metric $m$ ($e.g.$, $container\_spec\_cpu\_quota$) of collected metrics $M$ in Table \ref{tab:metric-labels}, PBScaler records a metric array $L_m$ for $v_i$ within a specified time window. PBScaler defines that $\textbf{\textit{P}}_{i, j}$ depends on the maximum value of the Pearson correlation coefficient between $L_t$ and $L_m$:

\begin{equation}
\textbf{\textit{P}}_{i, j} = \mathop{\max}\limits_{m\in M}(corr(L_t, L_m)).
\end{equation}

The Personalized PageRank algorithm determines the popularity of each node by randomly walking on the directed graph. However, some nodes may never point to others, causing the scores of all nodes to tend toward zero after many iterations of the random walk. To avoid falling into this "trap," a damping factor $\delta$ is applied, which allows the algorithm to jump out from these nodes according to a predefined rule. Typically $\delta$ is set to 0.15. The Personalized PageRank is represented as follows:

\begin{equation}
	\bm{v} = (1 - \delta)  \cdot \textbf{\textit{P}} \bm{v} + \delta  \cdot \bm{u},
\end{equation}
where $\bm{v}$ represents the probability that each microservice node is diagnosed as a PB. The preference vector $\bm{u}$  serves as the personalized rule to guide the algorithm to leap from the trap. The value of $\bm{u}$ is determined by the anomaly potential $\varphi$ of each node. The nodes with greater anomaly potential are preferred as starting points for the algorithm. The equation of the $k$-th iteration can be represented as:

\begin{equation}
	{\bm{v}}^{(k)} = (1 - \delta)  \cdot \textbf{\textit{P}} {\bm{v}}^{(k-1)} + \delta  \cdot \bm{u}.
\end{equation}
After multiple rounds of iterations,  $\bm{v}$ gradually converges. PBScaler then sorts the final results and produces the ranking list $\bm{rl}$. In Step 4 of Fig. \ref{fig:toporank}, the TopoRank algorithm undergoes multiple rounds of random walks. During this process, it reduces the suspicion levels on microservices A and C, assigning them low scores of 0.11 and 0.17, respectively. Conversely, microservice E is identified with the highest suspicion score of 0.37. The top-$k$ microservices (E and D) with the highest ranking list scores can be recognized as PBs. The whole process of TopoRank is depicted as Algorithm \ref{al:TopoRank}.

\subsection{Scaling Decision}

Given the PBs identified by the \textit{Performance Bottleneck Analysis}, the replicas for the PBs will be scaled to minimize the application's resource consumption while ensuring the end-to-end latency of microservices meets the SLO. Although abundant replicas can alleviate the performance degradation problem, they also consume a significant amount of resources. Consequently, it is essential to maintain a balance between performance guarantees and resource consumption. The process of \textit{Scaling Decision} will be modeled as a constrained optimization problem to achieve this balance. 

\subsubsection{Constrained Optimization Model}
\label{sec:Constrained_Optimization_Model}

The autoscaling optimization in our scenario seeks to identify an allocation schema that allocates a variable number of replicas for each PB. Given $n$ PBs that require scaling, we define a strategy as a set $\mathbb{X} = \{x_1, x_2, \cdots, x_n\}$, where $x_i$ denotes the number of replicas allocated for PB $i$. Before the optimization, the initial number of replicas for PBs can be expressed as $\mathbb{C} = \{c_1, c_2, \cdots, c_n\}$. It should be noted that the replicas constraint in PBScaler should be defined separately for scaled-down and scaled-up processes. During the scaled-up process, we limit the number of replicas for PBs as follows:

\begin{equation}
\begin{aligned}
\label{eq:constraint-scaled-up}
	s.t. \quad x_i &\geq c_i + 1 , \forall x_i \in \mathbb{X}, \forall c_i \in \mathbb{C},\\
 x_i &\leq c^{max}, \forall x_i \in \mathbb{X},
 \end{aligned}
\end{equation}
where $c^{max}$ represents the maximum number of replicas that a
microservice can scale to, given limited server resources. The constraint of the number of replicas during the scaled-down process can be expressed as:

\begin{equation}
\label{eq:constraint-scaled-down}
\begin{aligned}
	s.t. \quad x_i &\geq \max(c_i - \gamma, 1) , \forall x_i \in \mathbb{X}, \forall c_i \in \mathbb{C},\\
 x_i &\leq c_i, \forall x_i \in \mathbb{X},\forall c_i \in \mathbb{C}.
 \end{aligned}
\end{equation}
In Eq. (\ref{eq:constraint-scaled-down}), $\gamma$ (with the default value of two) denotes the maximum number of replicas reductions. This limit is reasonable since reducing the number of microservice replicas drastically can cause a short latency peak, as observed in experiments. 

The goal of the \textit{Scaling Descision} is to minimize the application’s resource consumption while maintaining its performance. Application performance is usually expressed by SLO violations that users are more concerned about. Therefore, the application performance reward can be detailed as:

\begin{equation}\label{eq:R1}
	R_1 = \left \{
	\begin{aligned}
		0,  &\quad SLO\; violation, \\
		1,  &\quad w/o \; SLO \; violation.
	\end{aligned}
	\right.
\end{equation}
During the  optimization process, the application's resource consumption, such as CPU and memory usage,  is unpredictable. To conservatively estimate resource consumption, we consider the ratio of PB replicas to the maximum number of allocatable replicas,  rather than calculating the cost of CPU and memory. We calculate the resource reward as:
\begin{equation}\label{eq:R2}
\begin{aligned}
	R_2 = 1 - \frac{\sum_{i=1} ^{n}{x_i}}{c^{max}\times n}.
	\end{aligned}
\end{equation}
Our objective is to guarantee performance while minimizing resource consumption. We leverage a weighted linear combination (WLC) method to balance the two objectives. The final optimization objective is defined as:
\begin{equation}\label{eq:R}
\begin{aligned}
	\max_{\mathbb{X}} (\lambda \cdot R_1(\mathbb{X}) + (1-\lambda) \cdot R_2(\mathbb{X})),
	\end{aligned}
\end{equation}
where $\lambda \in [0,1]$. We set $\lambda$ as a parameter to balance the application performance and resource consumption.

\subsubsection{SLO Violation Predictor}
To calculate the performance reward $R_1$, evaluating whether a strategy will cause the SLO violation in online applications is necessary.
A simple way is to execute candidate strategies directly in online applications and wait for feedback from the monitoring system. However, oscillations caused by frequent scaling in online applications will be inevitable. An alternative method is to train an evaluation model with historical metric data, which can simulate the feedback from online applications. Without interacting with online applications, this model predicts application performance based on the current application state. 

We use a vector $\bm{r}$ to denote the number of replicas for each microservice after executing the scaling strategy $\mathbb{X}$. $\bm{w}$ is a vector that denotes the current workload for each microservice. Because of the low time cost of bottleneck-aware optimization, it is reasonable to hypothesize that $\bm{w}$ will not change significantly during this period (see Section \ref{sec:PE}). Given the application state represented by workload $\bm{w}$ and replicas $\bm{r}$ of all microservices, an SLO violation predictor can be designed as:

\begin{equation}\label{eq:SLO-predictor}
	\psi(\bm{r}, \bm{w}) = \left \{
	\begin{aligned}
		0,  &\quad SLO\; violation, \\
		1,  &\quad w/o \; SLO \; violation,
	\end{aligned}
	\right.
\end{equation}
where $\psi$ is a binary classification model. Details of selecting an appropriate classification model will be discussed in Section \ref{sec:exp-validation}. The historical metric data used for training can be generated using either a classical scaling method (the Kubernetes autoscaler by default) or a stochastic method. We deployed an open-source microservice system on three nodes (with a total of 44 CPU cores and 220 GB of RAM) and performed elastic scaling. Prometheus gathered each microservice's workload and P90 tail latency at regular time intervals. By comparing the tail latency of the front-end microservice with the SLO, monitoring data for each time interval can be easily labeled.

\begin{figure}[t]
  \centering
  \includegraphics[width=\linewidth]{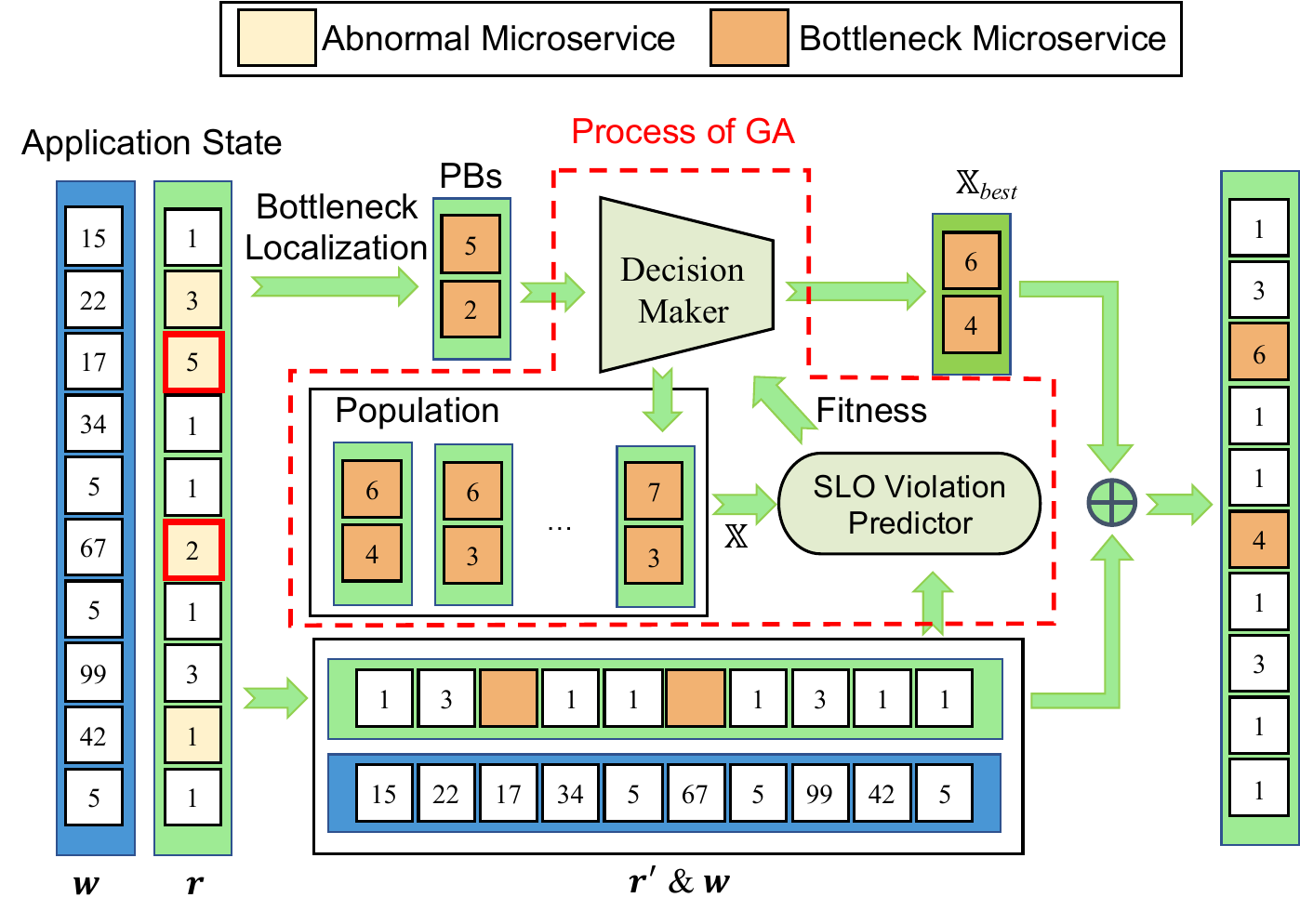}
  \caption{Illustration of the autoscaling optimization process.}
  \label{fig:GA_process}
\end{figure}

\subsubsection{Autoscaling Optimization}

As mentioned in Section \ref{sec:Constrained_Optimization_Model}, the tradeoff between performance and resource consumption can be modeled as a constrained optimization problem. To find a near-optimal strategy, PBScaler employs a genetic algorithm (GA) to generate and optimize scaling strategies that reduce resource consumption while meeting SLO requirements. By emulating natural selection in evolution, the GA improves the superior offspring while eliminating the inferior ones. Initially, the GA performs a random search to initialize a population of chromosomes in the strategy space, with each chromosome indicating a potential strategy for the optimization problem. Next, in each iteration, elite chromosomes with high fitness, called elites, will be selected for crossover or mutation to produce the next generation.

\begin{algorithm}[t]
	\renewcommand{\algorithmicrequire}{\textbf{Input:}}
	\renewcommand{\algorithmicensure}{\textbf{Output:}}
	\caption{GA-based Autoscaling Optimization}
	\begin{algorithmic}[1]
		\REQUIRE the number of iterations $I$ \\
		the size of population $N_p$ \\
  the size of elites $N_z$\\
  the probability of crossover $p_c$\\
  the probability of mutation $p_m$
		\ENSURE the best scaling strategy $s$
        \STATE $P_0$ $\leftarrow$ Randomly create a population with $N_p$ strategies
        \STATE Evaluate the fitness of each strategy in $P_{0}$
        \STATE $Z_0 \gets$ Get $N_z$ elites from $P_0$
        \STATE $s \gets $ Get the best strategy in $Z_0$
        \STATE $i \gets 0$
		\WHILE{$i < I$}
        
        \STATE $F_i \gets$ Set parents with $(N_p-N_z)$ strategies from $P_i$
        \STATE $O_i \gets$ Generate offspring by recombining $F_i$ with a probability of $p_c$  
        \STATE $O_i \gets$ Generate offspring by mutating $O_i$ with a probability of $p_m$
            
        \STATE $P_{i+1} \gets (Z_i \cup O_i)$
        \STATE Evaluate the fitness of each strategy in $P_{i+1}$
        \STATE $Z_{i+1} \gets$ Get $N_z$ elites from $P_{i+1}$
        \STATE $t \gets $ Get the best strategy in $Z_{i+1}$
        \STATE $f_s, f_t \gets$ Evaluate the fitness of $s$ and $t$
        \IF{$f_s < f_t$}
            \STATE $s \gets t$
        \ENDIF
        \STATE $i \gets i + 1$
        \ENDWHILE
	\end{algorithmic}
    \label{al:ga}
\end{algorithm}

The autoscaling optimization in our scenario seeks to identify a scaling strategy that allocates a variable number of replicas for each PB. The process of autoscaling optimization is illustrated in Fig. \ref{fig:GA_process}. In the beginning, PBScaler obtains each microservice's current number of replicas $\bm{r}$ and workload $\bm{w}$. After the \textit{Performance Bottleneck Analysis}, PBScaler identifies the PBs from $\bm{r}$ and filters out them to get $\bm{r^{\prime}}$. Then, the population of strategies for PBs is generated by the Decision Maker. Since the number of microservices to be scaled influences the speed and effect of the optimization algorithm (Section \ref{sec:exp-validation}), PBScaler assumes that only PBs need to be elastically scaled. In other words, the number of replicas in $\bm{r^{\prime}}$ will remain unchanged. The SLO violation predictor is responsible for evaluating the generated strategies. It should be noted that the strategy is merged with $\bm{r^{\prime}}$ and input to the SLO violation predictor together with $\bm{w}$. With the help of GA, the superior strategy $\mathbb{X}_{best}$ is selected and then merged with $\bm{r^{\prime}}$ to generate the final decision.



In the optimization phase, the Decision Maker generates and improves the scaling strategy for PB using the GA, as described in Algorithm \ref{al:ga}. After randomly generating a population within the strategy scope of each PB (Line 1), the Decision Maker estimates the fitness of each strategy based on Eq. (\ref{eq:R}) and stores the elites (Lines 2-3). In each iteration, the GA uses a tournament-based selection operator to pick out outstanding parents $F_i$ (Line 7). New offspring $O_i$ are generated through recombination and mutation (Lines 8-9) using a two-point crossover operator and a binary-chromosome mutation operator. By simulating natural selection, new offspring $O_i$ and elites $Z_i$ with higher fitness constitute a new population $P_{i+1}$ that enters the next iteration (Line 10). At each iteration, the current best strategy $t$ undergoes a fitness comparison with the historically best strategy $s$ (Lines 13-17). The best strategy $s$ after $I$ iterations will be considered as the value of final $\mathbb{X}_{best}$.


\section{EVALUATIONS}

In this section, we present the details of experimental scenarios for autoscaling, including a comparison of PBScaler with several state-of-the-art autoscaling algorithms from academia and industry.

\subsection{Experimental Setup}
\subsubsection{Microservice Platform}
Experiments were carried out in our private cloud cluster, consisting of three physical computers (one master node and two worker nodes) with a total of 44 Intel 2.40 GHz CPU cores and 220 GB of RAM. To evaluate autoscaling, we selected two open-source microservice applications as benchmarks: a) Online Boutique\footnote{https://github.com/GoogleCloudPlatform/microservices-demo}, a Web-based E-commerce demo application developed by Google. The system implements basic functions such as browsing products, adding items to shopping carts, and payment processing through the collaboration of ten stateless microservices and a Redis cache. b) Train Ticket\footnote{https://github.com/FudanSELab/train-ticket}:  a large-scale,  open-source microservice system developed by Fudan University. With more than 40 microservices and the usage of MongoDB and MySQL for data storage, Train Ticket can satisfy a variety of workflows, such as online ticket browsing,  booking, and purchasing. Because of cluster resource constraints, we limited each microservice to no more than eight replicas. The source code is available on Github\footnote{https://github.com/WHU-AISE/PBScaler}.

\begin{figure}[t]
\centering{
 \subfigure[ Wiki ]{
 \includegraphics[width=0.31\linewidth]{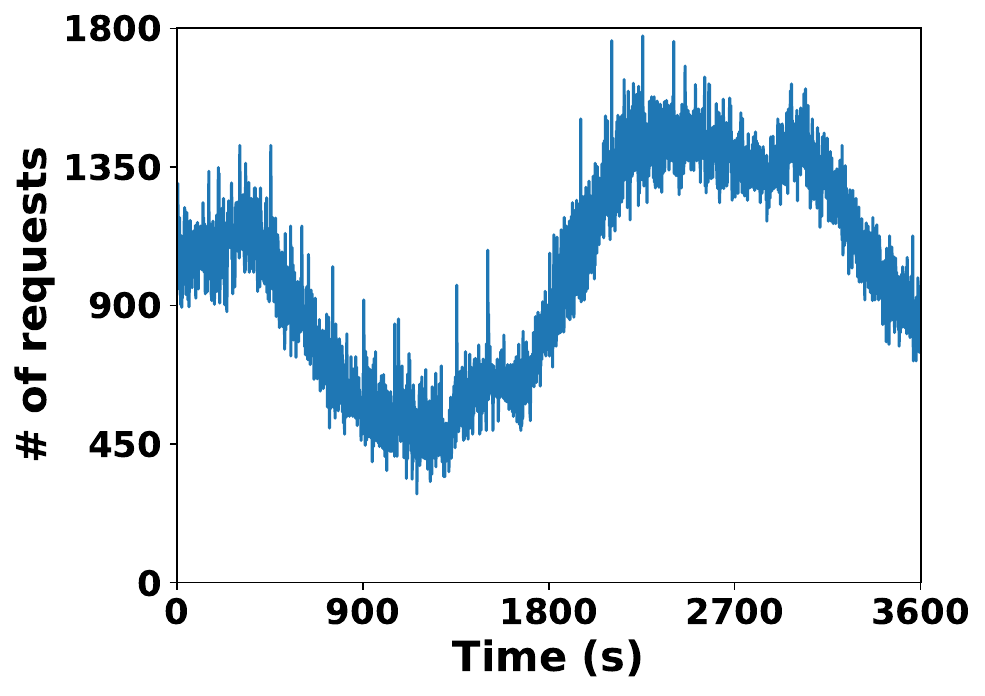} } 
 \subfigure[ EW1 ]{  \includegraphics[width=0.31\linewidth]{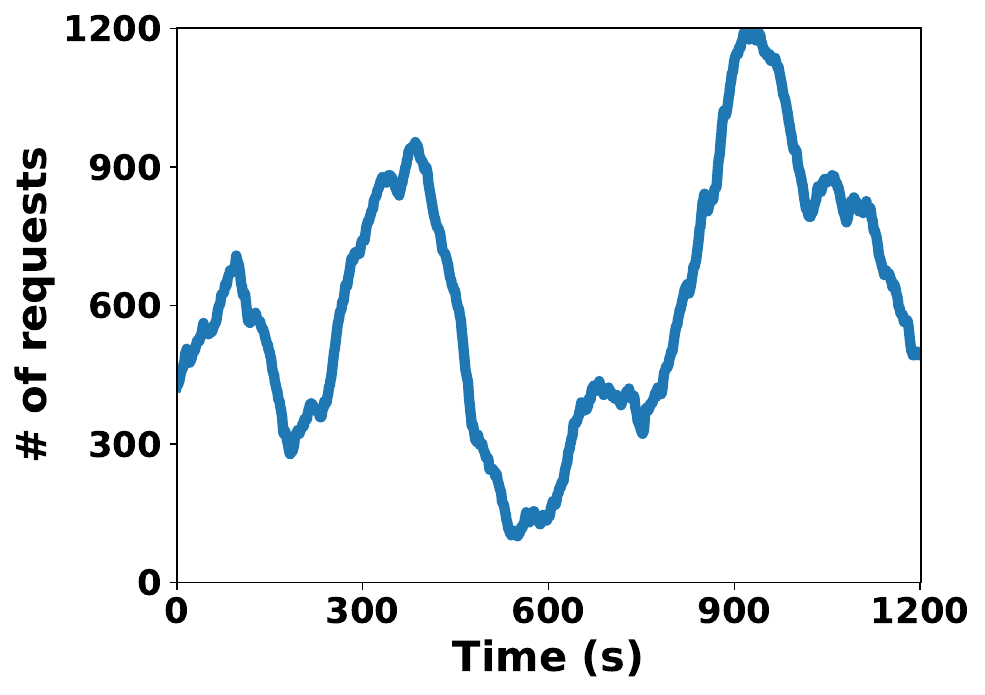}}
 \subfigure[ EW2 ]{  \includegraphics[width=0.31\linewidth]{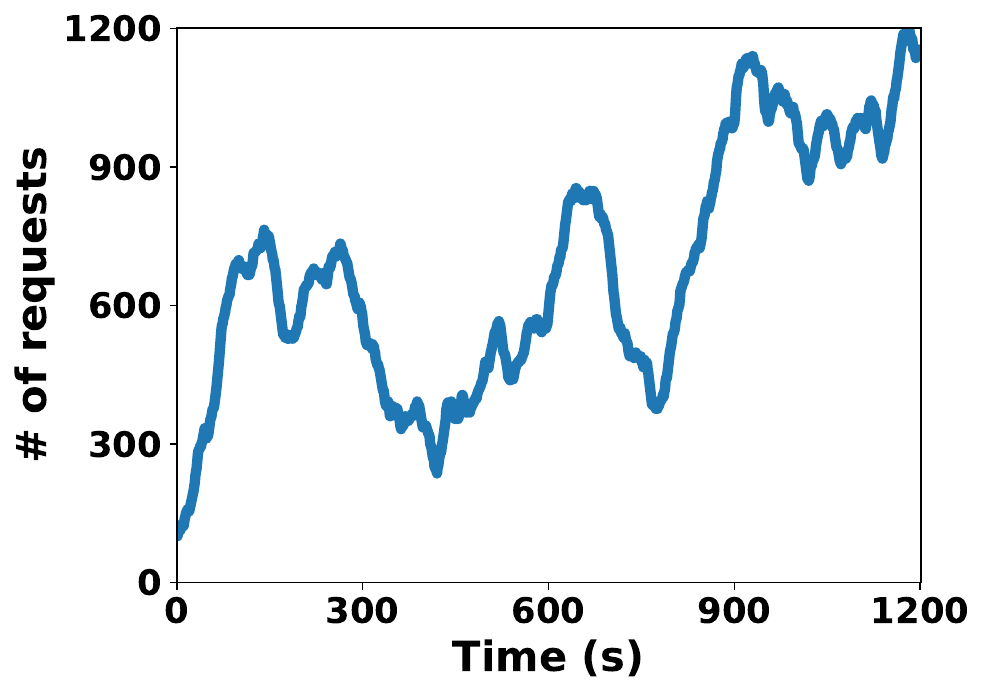}}
 \subfigure[ EW3 ]{  \includegraphics[width=0.31\linewidth]{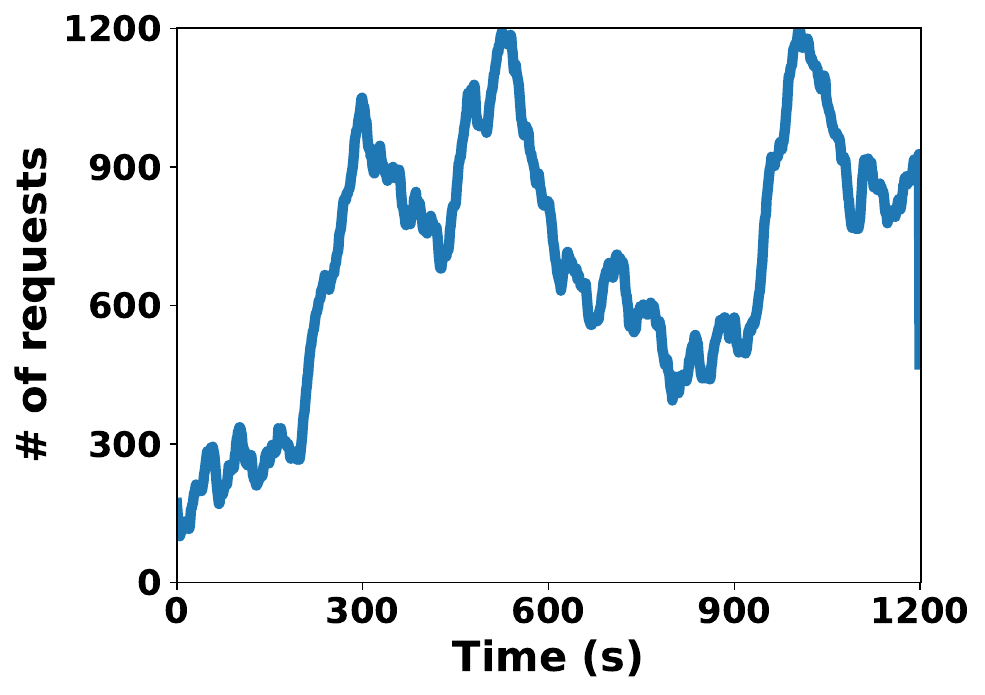}}
 \subfigure[ EW4 ]{  \includegraphics[width=0.31\linewidth]{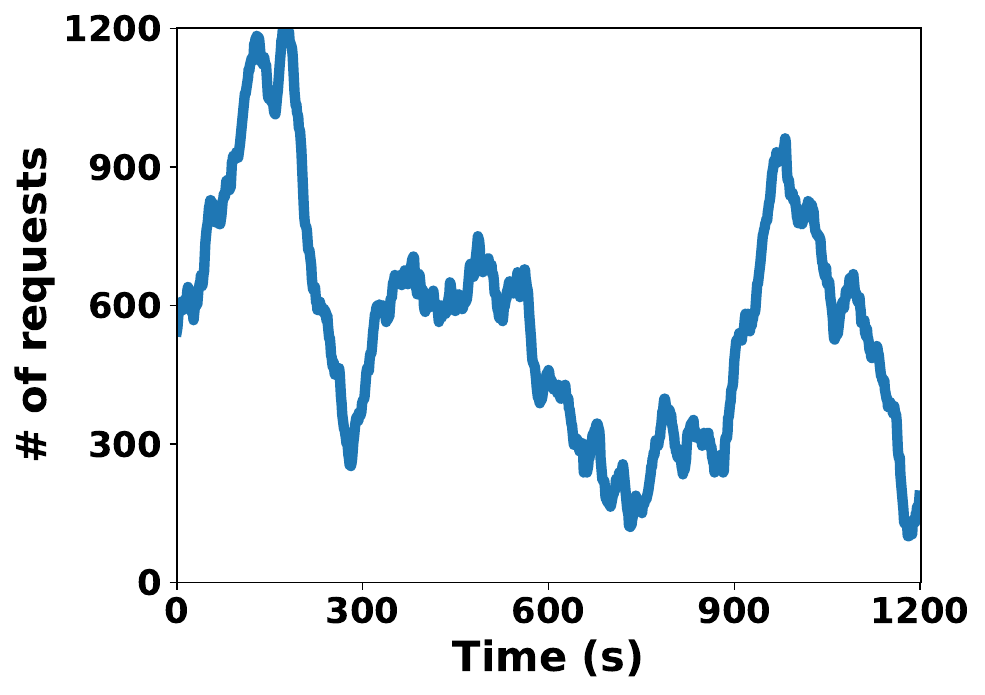}}
 \subfigure[ EW5 ]{  \includegraphics[width=0.31\linewidth]{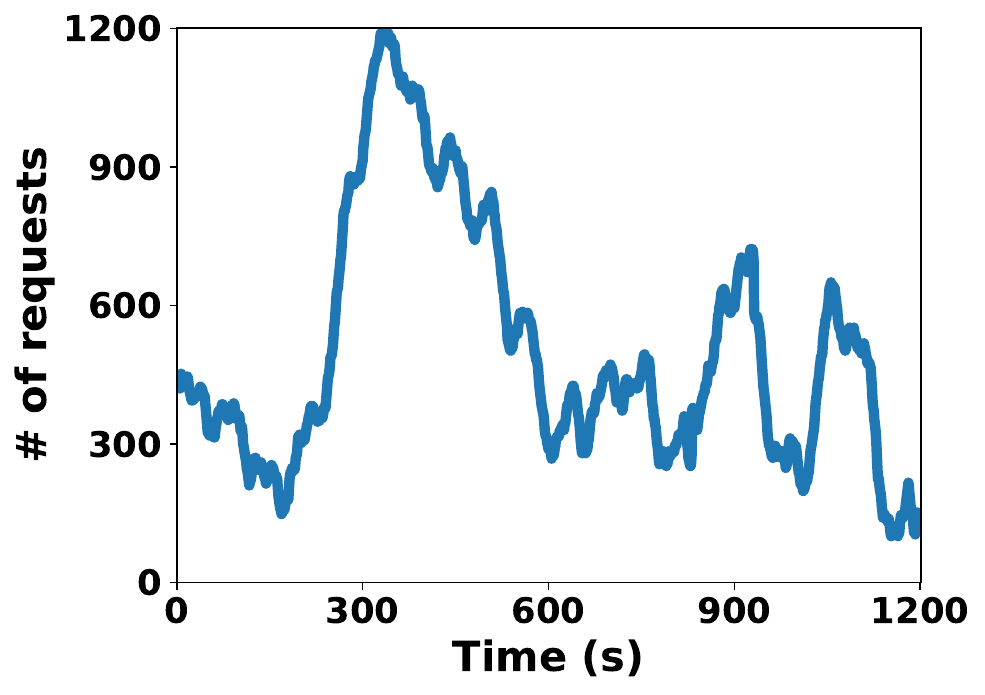}}
}
\caption{The fluctuation of one-hour real-world workloads (Wiki) and twenty minutes emulated workloads (EW).}
\vspace{-4mm}
\label{fig:workloads}
\end{figure}

\subsubsection{Workload}
We evaluated the effectiveness of PBScaler under various traffic scenarios, using a real-world Wikipedia workload from the Wiki-Pageviews \cite{wiki-dataset} on March 16, 2015, and five emulated workloads (EW1 $ \sim $ EW5), inspired by the experiments conducted by Abdullah \textit{et al.} in \cite{abdullah2020burst}. We compressed the real-world workload to one hour and scaled it to an appropriate level for our cluster. The five emulated workloads exhibited various patterns, such as single peak, multiple peaks, rising, and dropping, and were limited to a duration of twenty minutes. Fig. \ref{fig:workloads} depicts the fluctuation of these workloads.

\subsubsection{Baseline Methods}
We compare PBScaler with several state-of-the-art microservice autoscaling methods from academia and industry, which perform dynamic horizontal scaling of microservices from the perspectives of static thresholds, control theory, and black-box optimization. 

\begin{itemize}
\item \textbf{Kubernetes Horizontal Pod Autoscaling (KHPA)}: It is the default horizontal scaling scheme of Kubernetes. By customizing a threshold $T$ for a certain resource $R$ (CPU usage as the default) and aggregating the resource usage $U_i^R$ from all replicas of a microservice, KHPA defines the target number of replicas as $n = \lceil \sum_{i \in ActivePods}{U_i^R} \, / \, T  \rceil$. 

\item \textbf{MicroScaler} \cite{yu2019microscaler}: It is an autoscaling tool that uses a black-box optimization algorithm to determine the optimal number of replicas for a microservice. MicroScaler calculates the microservice's P90/P50 for classification and then performs four iterations of Bayesian Optimization to make a scaling decision. 
\item \textbf{SHOWAR} \cite{baarzi2021showar}: It is a hybrid autoscaling technology. We reproduced the horizontal scaling part in SHOWAR, which uses the PID control theory to gradually bring the observed metric close to the user-specified threshold. In our implementation, we replaced the run queue latency with the more common P90 latency since the former requires an additional eBPF tool.

\end{itemize}

\definecolor{mygray}{gray}{.9}
\newcommand{\red}[1]{\textcolor{red}{#1}}
\newcommand{\grey}[1]{\textcolor{gray}{#1}}
\begin{table*}[t]
\small
\caption{Performance of state-of-the-art methods and PBScaler under real and emulated workloads. SLO violation rate and cost are reported.}
\label{tab:all-performance}

\setlength{\tabcolsep}{3mm}{
\begin{tabular}{l||cccccc||cccccc}
\hline
\multicolumn{1}{c||}{\multirow{2}{*}{Methods}} & \multicolumn{6}{c||}{Online Boutique} & \multicolumn{6}{c}{Train Ticket} \\ \cline{2-13} 
\multicolumn{1}{c||}{} & \multicolumn{1}{c}{Wiki} & \multicolumn{1}{c}{EW1} & \multicolumn{1}{l}{EW2} & \multicolumn{1}{c}{EW3} & \multicolumn{1}{l}{EW4} & \multicolumn{1}{c||}{EW5} & \multicolumn{1}{c}{Wiki} & \multicolumn{1}{c}{EW1} & \multicolumn{1}{l}{EW2} & \multicolumn{1}{c}{EW3} & \multicolumn{1}{l}{EW4} & \multicolumn{1}{c}{EW5} \\ \cline{1-1}
\hline
\rowcolor{mygray}\multicolumn{13}{l}{SLO violation rate (\%)} \\

\hline
None & 79.64 & 57.26 & 67.22 & 70.12 & 56.85 & 39.00 & 94.87 & 48.37 & 53.11 & 57.54 & 35.88 & 37.89 \\
KHPA & \textbf{4.86} & 13.45 & 37.08 & 31.95 & 20.00 & 25.10 & 10.48 & 37.50 & 40.09 & 30.46 & 29.72 & 30.90 \\
MicroScaler & 8.18 & 17.43 & 17.45 & 31.54 & 22.41 & 18.26 & 46.70 & 29.61 & 30.48 & 47.57 & 50.29 & 42.16 \\
SHOWAR & 13.74 & 13.33 & \textbf{10.92} & 27.39 & 12.08 & 25.00 & 9.39 & 20.14 & 22.92 & 19.23 & 25.66 & 21.04 \\
PBScaler & 5.69 & \textbf{7.88} & 12.45 & \textbf{8.30} & \textbf{11.62} & \textbf{8.71} & \textbf{5.62} & \textbf{15.00} & \textbf{19.40} & \textbf{18.14} & \textbf{16.24} & \textbf{14.22} \\
\hline
\rowcolor{mygray}\multicolumn{13}{l}{Cost (\$)} \\
\hline
None & 1.93 & 0.50 & 0.62 & 0.65 & 0.61 & 0.64 & 11.62 & 3.03 & 3.06 & 3.25 & 3.15 & 3.39 \\
KHPA & 3.48 & 1.03 & 1.06 & 1.08 & 1.11 & 1.09 & 18.29 & 5.50 & 5.71 & 5.66 & 5.84 & 5.27 \\
MicroScaler & 3.09 & 0.79 & 0.72 & \textbf{0.71} & 0.76 & \textbf{0.68} & 15.44 & 3.76 & 3.52 & 4.70 & 4.86 & 4.46 \\
SHOWAR & \textbf{2.49} & 0.71 & 0.83 & 0.85 & 0.79 & 0.69 & \textbf{13.64} & 3.86 & 3.82 & 4.21 & 4.09 & 3.56 \\
PBScaler & 2.57 & \textbf{0.68} & \textbf{0.69} & 0.72 & \textbf{0.63} & 0.69 & 14.02 & \textbf{3.57} & \textbf{3.30} & \textbf{3.64} & \textbf{3.69} & \textbf{3.14}\\
\hline
\end{tabular}}
\end{table*}

\begin{figure*}[t]
\centering{
\subfigure[ EW1 ]{
\includegraphics[width=0.3\linewidth]{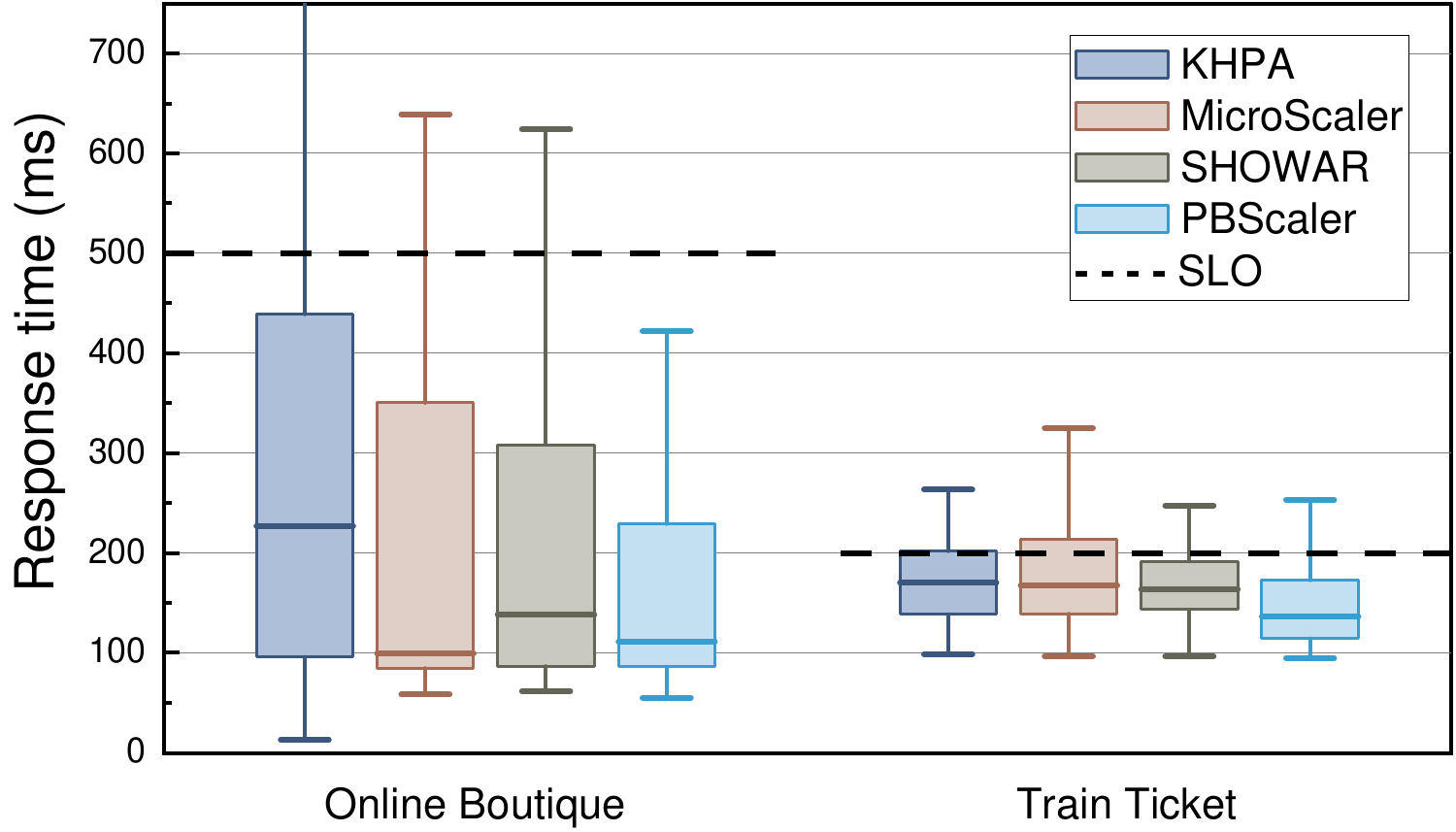}
}
\subfigure[ EW2 ]{
\includegraphics[width=0.3\linewidth]{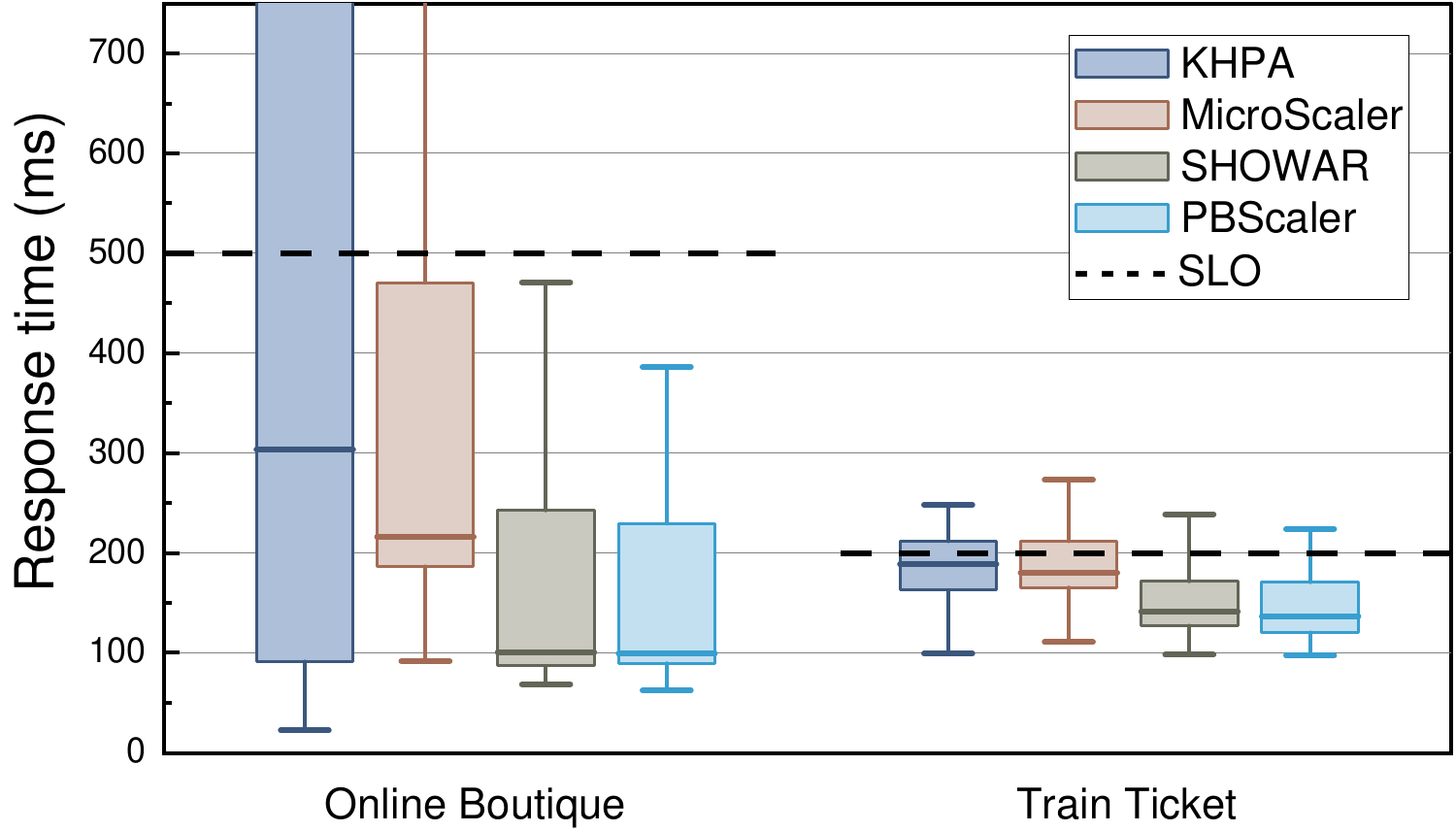}
}
\subfigure[ EW3 ]{
\includegraphics[width=0.3\linewidth]{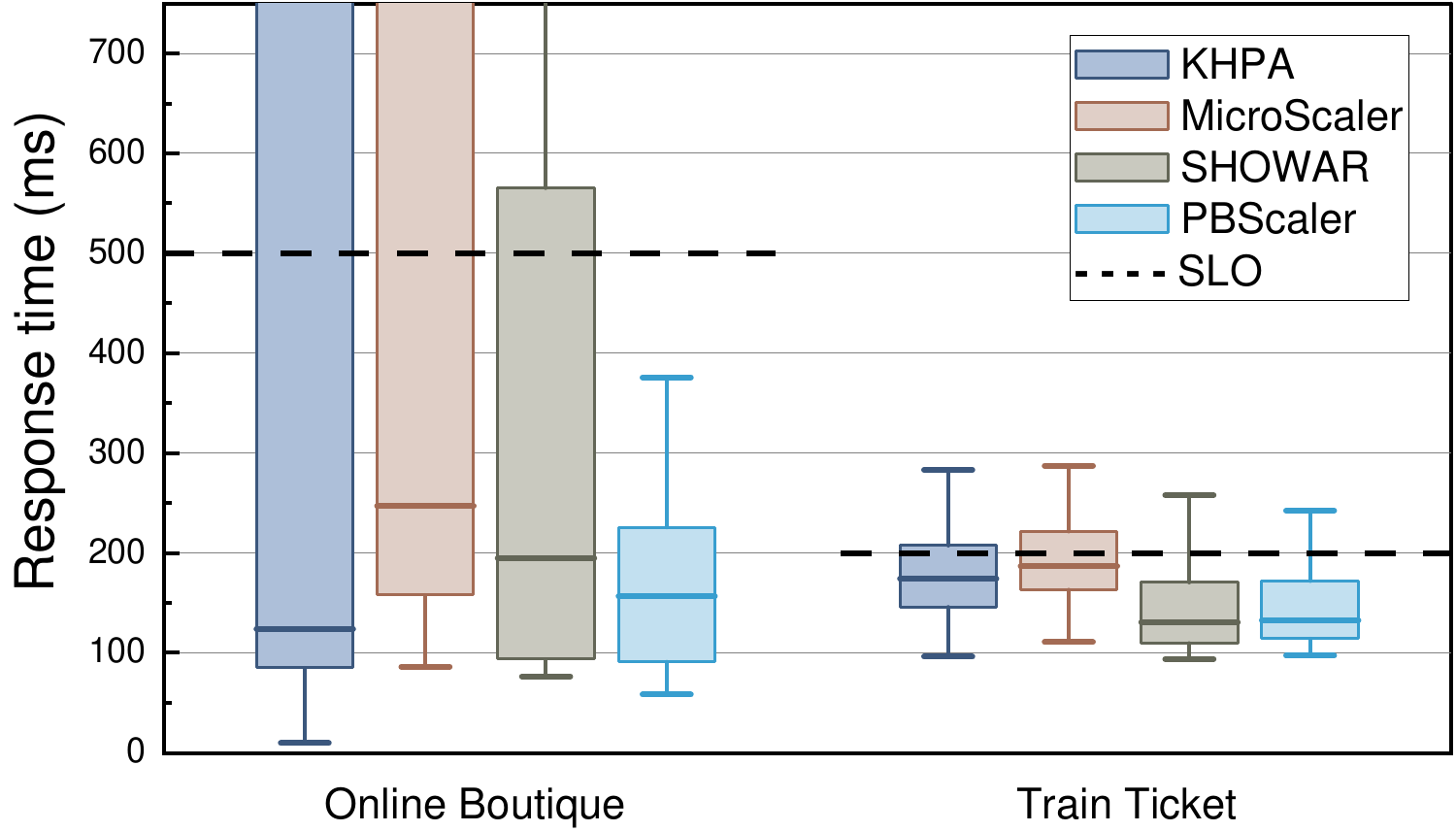}
}
\subfigure[ EW4 ]{
\includegraphics[width=0.3\linewidth]{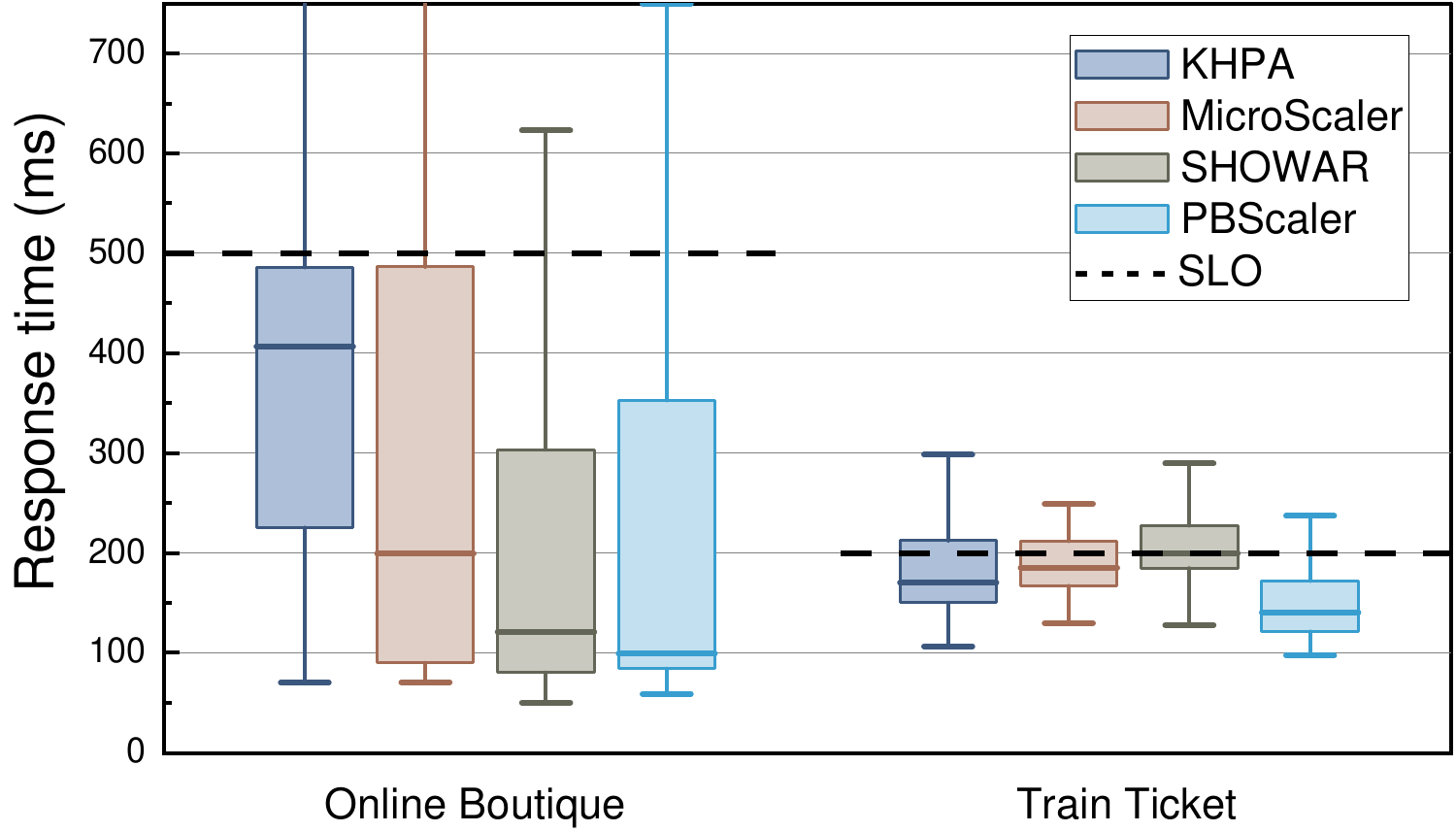}
}
\subfigure[ EW5 ]{
\includegraphics[width=0.3\linewidth]{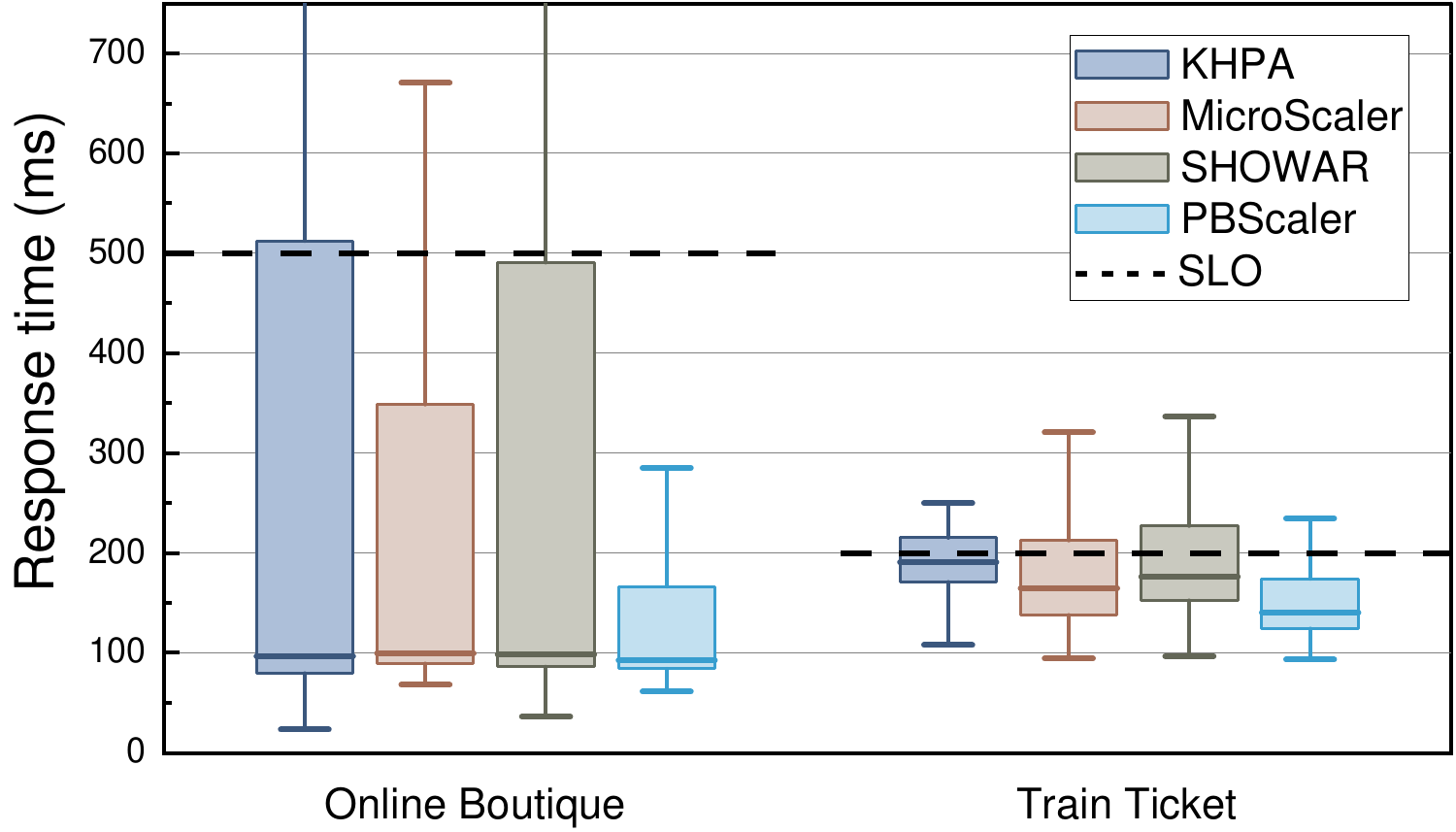}
}
\subfigure[ Wiki ]{
\includegraphics[width=0.3\linewidth]{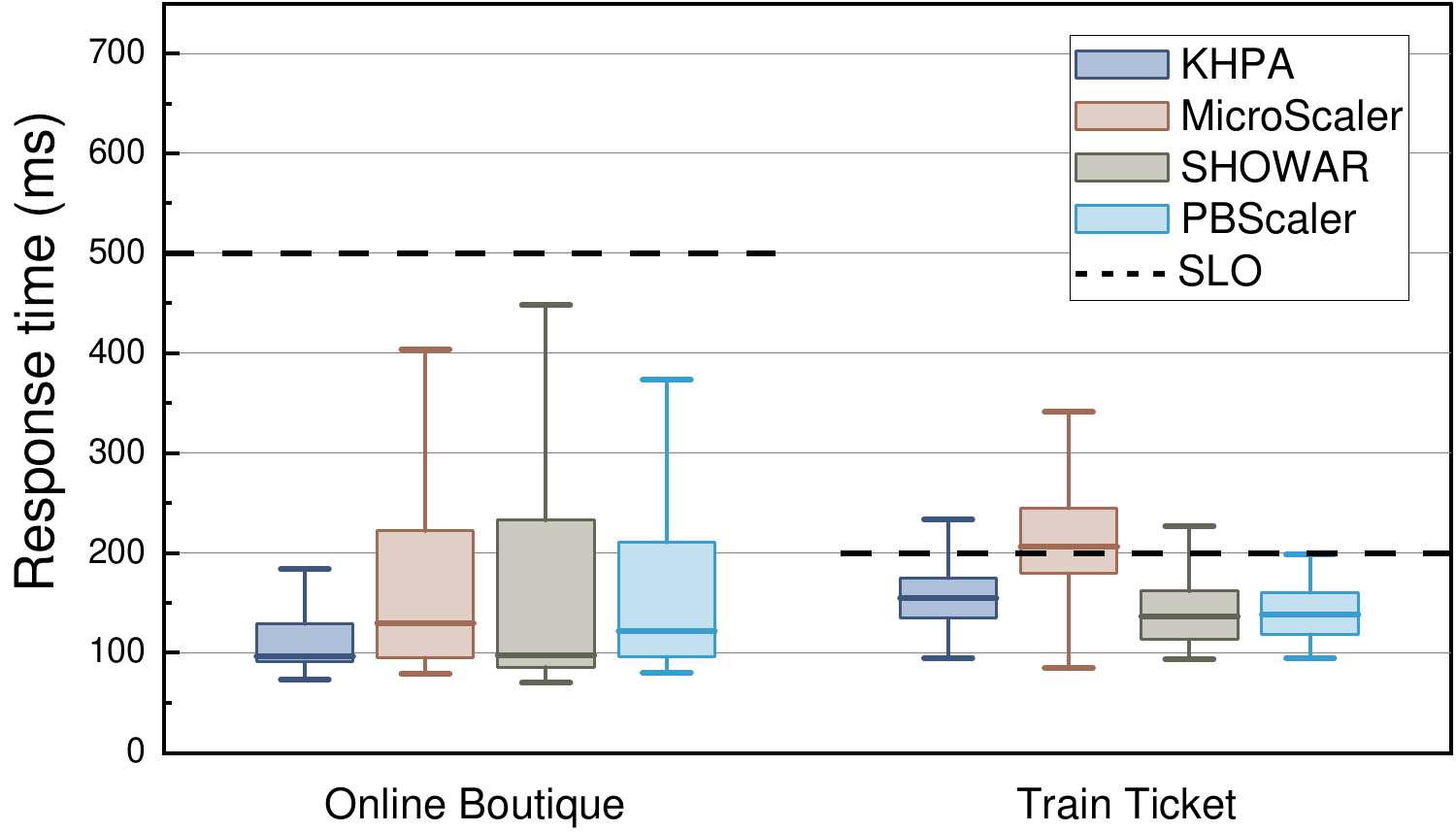}
}
}
\caption{The latency distribution for different methods under real and emulated workloads.}
\label{fig:latency-distribution}
\end{figure*}

\subsubsection{Experimental Parameters and Evaluation Criteria}
In our experiments, we fixed the collection interval of Prometheus to five seconds. With the increase in experiment time and workloads, the data volume required by stateful microservices like MongoDB will also grow. Eventually, the data volume will exceed the available memory, necessitating the use of disk storage. This transition can cause performance degradation that cannot be remedied through autoscaling. Hence, we limit workload testing to stateless traces. The SLO values for the Online Boutique and the Train Ticket were set to 500 ms and 200 ms, respectively. In the SLO violation detection and redundancy checking module, PBScaler first sets the action boundary $\alpha$ to 0.2 to reduce noise interference, as done in SHOWAR. Then, we 
empirically set the degree of significance $\beta$ to 0.9 to control the workload level that triggers scaling. For bottleneck localization, the impact factor $\sigma$ of the topological potential is set to 1, and the top-$k$ ($k$ =2) microservices with the highest score in $\bm{rl}$ will be considered PBs.

We choose the SLO violation rate, resource consumption, and response time to evaluate the performance of autoscaling methods. An autoscaling approach is considered more effective if it can reduce response time, SLO violation rate, and resource consumption. We define the SLO violation rate as the percentage of the end-to-end P90 tail latency that exceeds the SLO. Resource consumption is calculated following the method presented in \cite{ding2021copa}, where the CPU price is 0.00003334\$ (vCPU/s) and the memory price is 0.00001389\$ (G/s). The total resource consumption is obtained by summing the cost of memory and CPU.

\begin{table}[t]\tiny
\caption{Time cost of four modules in PBScaler.}
\label{tab:time-cost}
\resizebox{\linewidth}{!}{ 
\begin{tabular}{l||c||c}
\hline
\multicolumn{1}{c||}{\textbf{Modules}} & \textbf{\begin{tabular}[c]{@{}c@{}}Online \\ Boutique\end{tabular}} & \textbf{\begin{tabular}[c]{@{}c@{}}Train \\ Ticket\end{tabular}} \\
\hline
SLO Violation Detection & 0.29s & 1.03s \\
Redundancy Checking & 0.11s & 0.15s \\
PBA & 0.79s & 3.1s \\
Decision Maker & 3.36s & 3.58s \\
\hline
\end{tabular}
}
\vspace{-2mm}
\end{table}

\subsection{Performance Evaluation}
\label{sec:PE}

Table \ref{tab:all-performance} compares the SLO violation rates and resource costs for the four autoscaling methods in two microservice applications with different workloads. The None method is used as a reference and performs no autoscaling operation. Its results are presented in grey and are excluded from the comparison. 

In general, PBScaler outperforms the competing approaches in reducing SLO violations and minimizing resource overhead under six workloads in both microservices systems. In particular, the SLO violation rate of PBScaler in Train Ticket is,  on average,  4.96\% lower than that of the baseline methods, while the resource cost is reduced by an average of 0.24\$. These results show that PBScaler can perform elastic scaling for bottleneck microservices in large-scale microservice systems quickly and precisely, thereby reducing SLO violations and saving resources. Regarding the six workloads in Online Boutique, PBScaler also achieves the lowest SLO violations in four of them and minimizes resource consumption in three emulated workloads.



Fig. \ref{fig:latency-distribution} depicts the box plots of latency distribution for different methods under six workloads,  exploring the impact of each method on the performance of the microservice system. It can be seen that the majority of the autoscaling methods can keep the median of the latency distribution below the red dotted line (SLO). However, only PBScaler goes a step further to reduce the third quartile significantly below the SLO for all workloads.

To evaluate the time cost of using PBScaler for elastic scaling, the average time required by each module in PBScaler is collected and counted. As reported in Table \ref{tab:time-cost}, the total time cost of all PBScaler modules in Online Boutique is less than one monitoring interval (i.e., 5s), while the same metric for Train Ticket is less than two monitoring intervals. Thanks to the PBA that narrows the decision-making scope, the time cost of the Decision Maker does not increase much (no more than 6.6\%) when the application is switched from Online Boutique to Train Ticket, despite the increased number of microservices.  However, we recognize the limitation that the time consumption of PBA quickly rises as the microservice scale grows, which will be our future work.





\begin{figure}[t]
  \centering
  \includegraphics[width=0.9\linewidth]{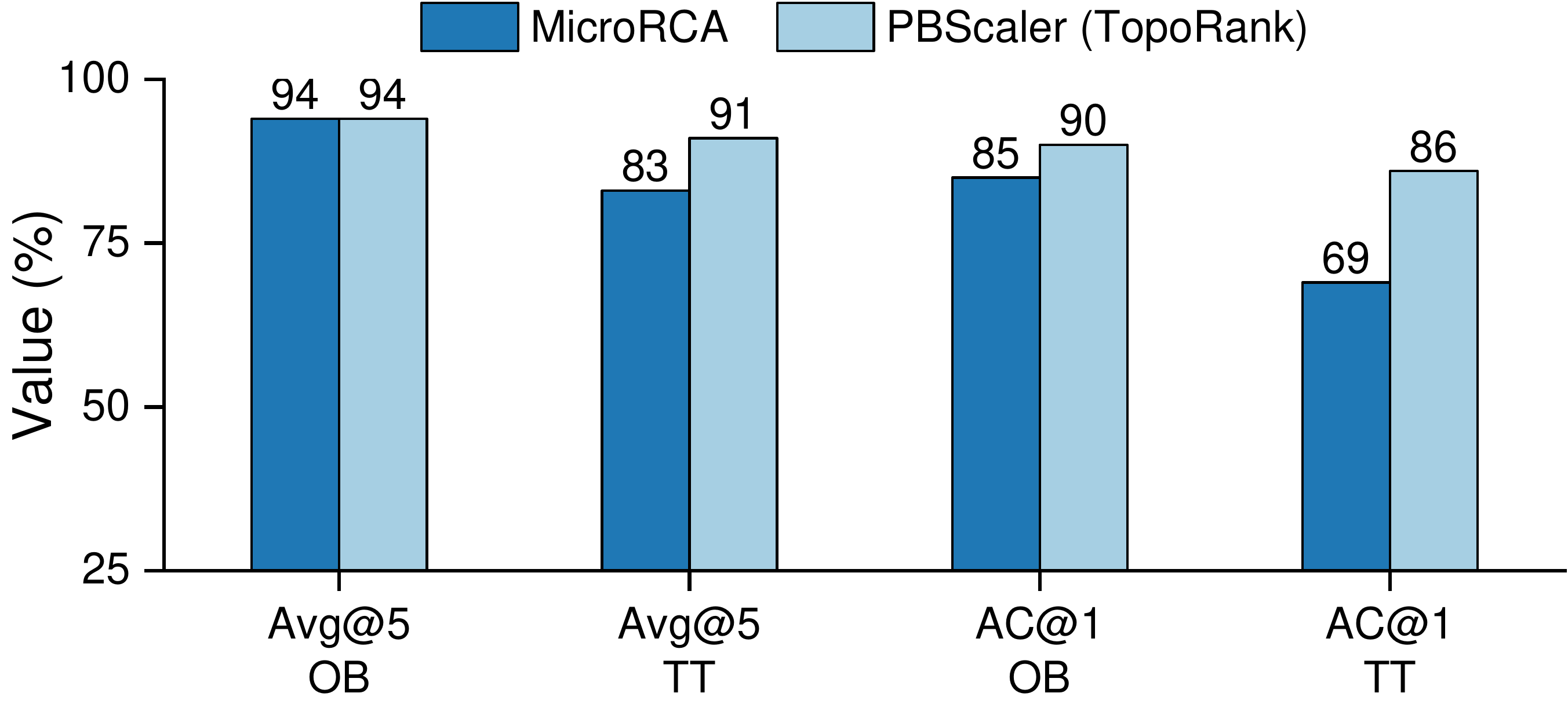}
  \caption{Performance comparison of bottleneck localization on Online Boutique(OB) and Train Ticket(TT).}
  \label{fig:PBA-performance}
\end{figure}

\begin{figure}[t]
\centering{
\subfigure[Online Boutique]{\includegraphics[width=0.48\linewidth]{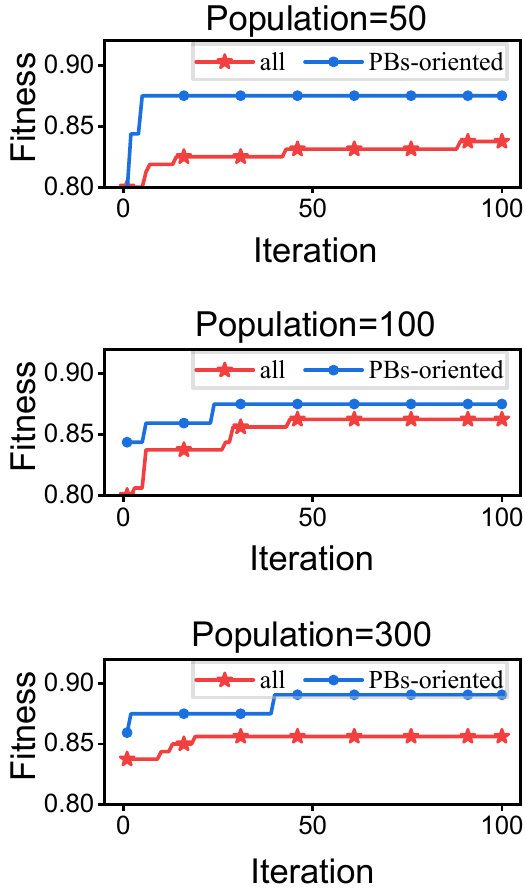} } 
\subfigure[Train Ticket]{\includegraphics[width=0.48\linewidth]{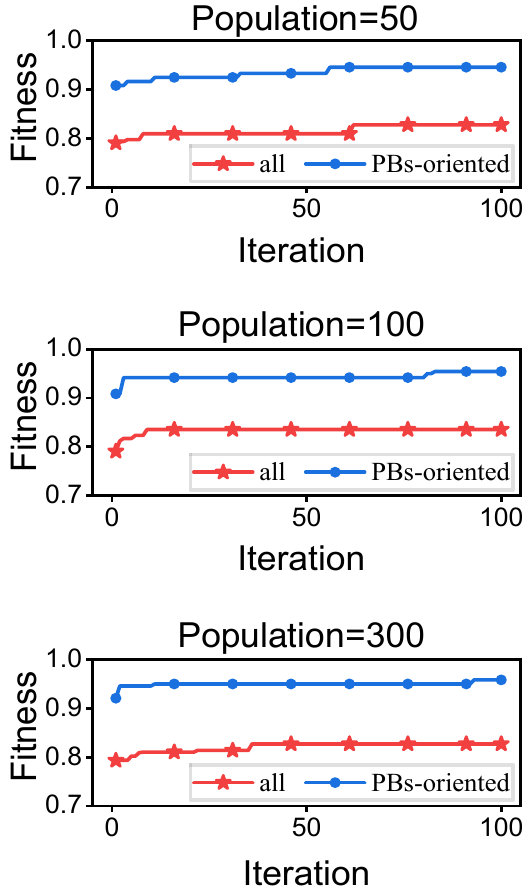} } 
}
\caption{The iterative process of the genetic algorithm under Online Boutique and Train Ticket.}
\label{fig:fitness-iteration-train}
\end{figure}

\subsection{Effectiveness Analysis of Components}
\label{sec:exp-validation}

\subsubsection{Performance Comparision of Bottleneck Localization}
To evaluate whether the TopoRank algorithm can effectively locate PBs caused by burst workloads, we injected exceptions, such as CPU overload, memory overflow, and network congestion, into Online Boutique and Train Ticket through Chaos Mesh. These exceptions are typically caused by high-workload conditions. The TopoRank algorithm was used to analyze the metrics and identify performance bottlenecks for these exceptions. The localization results were then compared to MicroRCA \cite{wu2020microrca}, a baseline method for microservice root cause analysis. $AC@k$ measures the accuracy of the real PBs in the top-$k$ results, and $Avg@k$ is the average accuracy at the top-$k$ results. These metrics can be calculated as follows.

\vspace{-2mm}
\begin{equation}\label{eq:PR@N}
	AC@k = \frac{1}{\lvert A \rvert} \sum \limits_{a \in A} { \frac{\lvert RT@k \cap PBs \rvert}{min(k, \lvert PBs \rvert)}},
\end{equation}

\begin{equation}\label{eq:MAP}
	Avg@k = \frac{1}{\lvert A \rvert} \sum \limits_{a \in A} {\sum \limits_{k =1}^{\lvert A \rvert} AC@k},
\end{equation}
where $A$ represents the set of exceptions, and $RT@k$ refers to the top-$k$ microservices in the ranking list. Fig. \ref{fig:PBA-performance} presents the $AC@1$ and $Avg@5$ values of TopoRank and MicroRCA across different microservice applications. The results indicate that TopoRank performs better than MicroRCA in both metrics. This is primarily due to the fact that TopoRank takes into account both the anomaly potential and microservice dependencies when performing Personalized PageRank.

The primary purpose of bottleneck location is to narrow down the strategy space and expedite the discovery of the optimal strategy. We perform GA iterations on both PBs and all microservices to demonstrate the influence of bottleneck localization on optimization. Fig. \ref{fig:fitness-iteration-train} depicts the iterative process under the microservice systems and demonstrates that as the population increases, the PB-aware strategy significantly outperforms the approach that scales for all microservices in terms of fitness. The PB-aware strategy can obtain superior fitness in less than five iterations. In contrast, the all-microservices-involved method requires larger populations and more iterations to achieve the same level of fitness. This is attributed to the fact that the PB-aware strategy aids the genetic algorithm in reducing the optimization range precisely and accelerating the acquisition of superior solutions.

\begin{table}[t]
\caption{Precision and Recall of four ML methods for SLO violation prediction}
\label{tab:PR&Recall}
\centering
\begin{tabular}{l|cc|cc}
\Xhline{0.8pt}
\multirow{2}{*}{Method} & \multicolumn{2}{c|}{Train Ticket} & \multicolumn{2}{c}{Online Boutique} \\ \cline{2-5} 
 & $Precision$ & $Recall$ & $Precision$ & $Recall$ \\ \hline
SVM & 0.819 & 0.961 & 0.865 & 0.915 \\
Decision Tree & 0.891 & 0.918 & 0.927 & 0.941 \\
Random Forest & \textbf{0.919} & \textbf{0.963} & \textbf{0.956} & \textbf{0.969} \\
MLP & 0.799 & 0.930 & 0.831 & 0.907 \\ \Xhline{0.8pt}
\end{tabular}
\end{table}

\begin{figure}[t]
\centering{
 \subfigure[ Replicas fluctuation]{\includegraphics[width=0.48\linewidth]{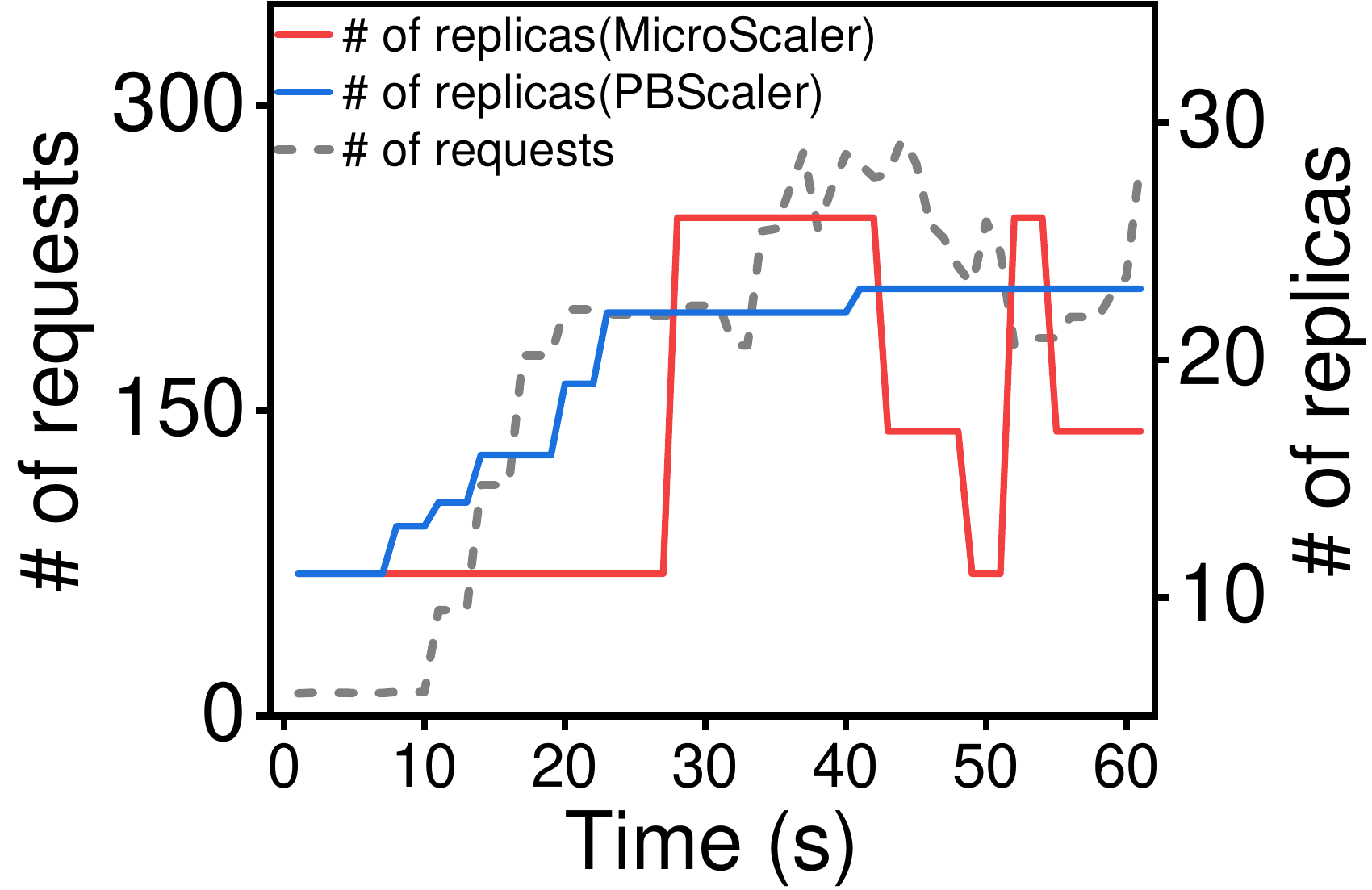} } 
 \subfigure[ Latency fluctuation]{  \includegraphics[width=0.48\linewidth]{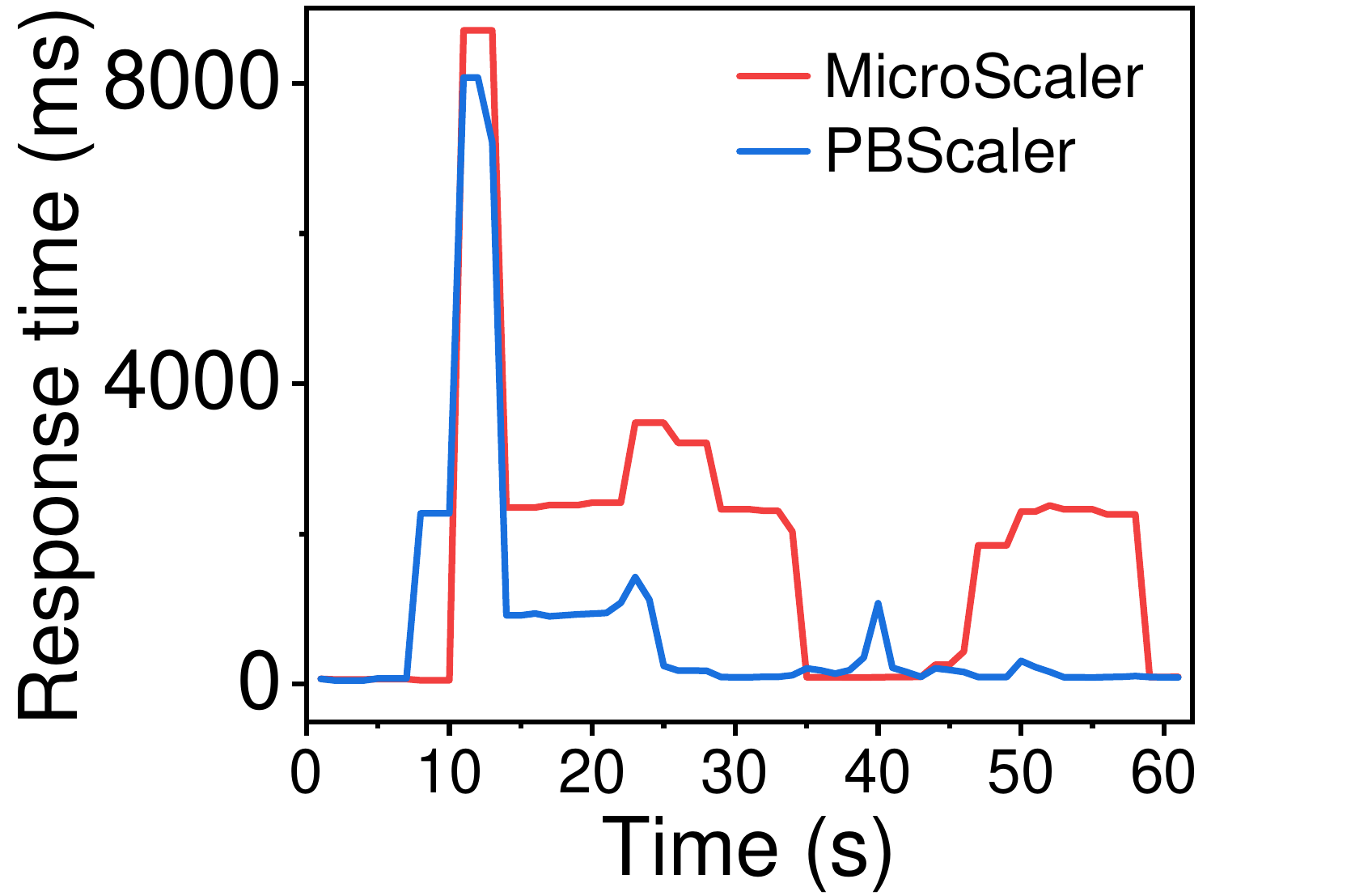}}
}
\caption{The replica and latency fluctuation of PBScaler and MicroScaler under the Wiki workload.}
\label{fig:burst-wave-comparision}
\end{figure}

\subsubsection{Effectiveness of the SLO Violation Predictor}
The objective of the SLO violation predictor is to forecast the result of the optimization strategy directly rather than waiting for feedback from the online application. We determine whether performance issues will occur based on the number of replicas and workloads of each microservice. Selecting a suitable binary classification model for the prediction task is critical. With a data collection interval of five seconds, we collected two datasets, including a 3.1k historical sampling dataset for Train Ticket and a 1.5k dataset for Online Boutique, in our cluster. For training and testing on these two datasets, we adopt four classical machine learning (ML) methods, including Support Vector Machine (SVM), Random Forest, Multilayer Perceptron (MLP), and Decision Tree. We employ typical binary classification indicators to assess the performance of experimental methods: $Precision = \frac{TP}{TP + FP}$ and $Recall = \frac{TP}{TP + FN}$, where $TP$ represents the number of identified SLO violations; $FN$ and $FP$ represent the number of unrecognized SLO violations and the number of false-positive SLO violations, respectively. Table. \ref{tab:PR&Recall} shows the performance comparison of the four methods for SLO violation prediction. According to the effects of the two datasets, we finally choose Random Forest as the primary algorithm for the SLO violation predictor.

To demonstrate that the SLO violation predictor can substitute the feedback from the real environment, we compare PBScaler, which employs the SLO violation predictor, with MicroScaler, which collects feedback from the online system. We injected burst workloads into the Online Boutique and made only one microservice abnormal to eliminate the difference in bottleneck localization between the two methods. As shown in Fig. \ref{fig:burst-wave-comparision}, with the guidance of the predictor, the number and the frequency of decision-making attempts made by PBScaler are much lower than those of MicroScaler. Reducing online attempts in a cluster will evidently reduce the risk of oscillations.

\section{CONCLUSIONS}
This paper presents PBScaler, a bottleneck-aware autoscaling framework designed to prevent performance degeneration in microservice-based applications. PBScaler collects real-time performance metrics of applications using the service mesh technology and dynamically builds a correlation graph among microservices. To handle abnormal microservices caused by external dynamic workloads and intricate invocations among microservices, PBScaler employs TopoRank, a random walk algorithm based on the topological potential theory, to identify bottleneck microservices. Furthermore, PBScaler performs an offline evolutionary algorithm to optimize scaling strategies guided by an SLO violation predictor. Experimental results indicate that PBScaler can minimize resource consumption while achieving lower SLO violations. 

In the future, we plan to improve our work from the following two aspects. Firstly, we will explore the potential of using bottleneck awareness in finer-grained resource (e.g., CPU and memory) management. Secondly, we will explore how to circumvent the interference of stateful microservices in autoscaling since the performance degradation from stateful microservices may disrupt the autoscaling controller. Thirdly, we will improve the efficiency of performance bottleneck analysis for large-scale microservice systems.


\section*{Acknowledgement}

This work is supported by the National Key Research and Development Program of China (No. 2022YFF0902701) and the National Natural Science Foundation of China (Nos. 62032016, 61832014, and 61972292).

\ifCLASSOPTIONcaptionsoff
  \newpage
\fi

\bibliographystyle{IEEEtran}
\bibliography{PBScaler-base}


\begin{IEEEbiography}[{\includegraphics[width=1in,height=1.25in,clip,keepaspectratio]{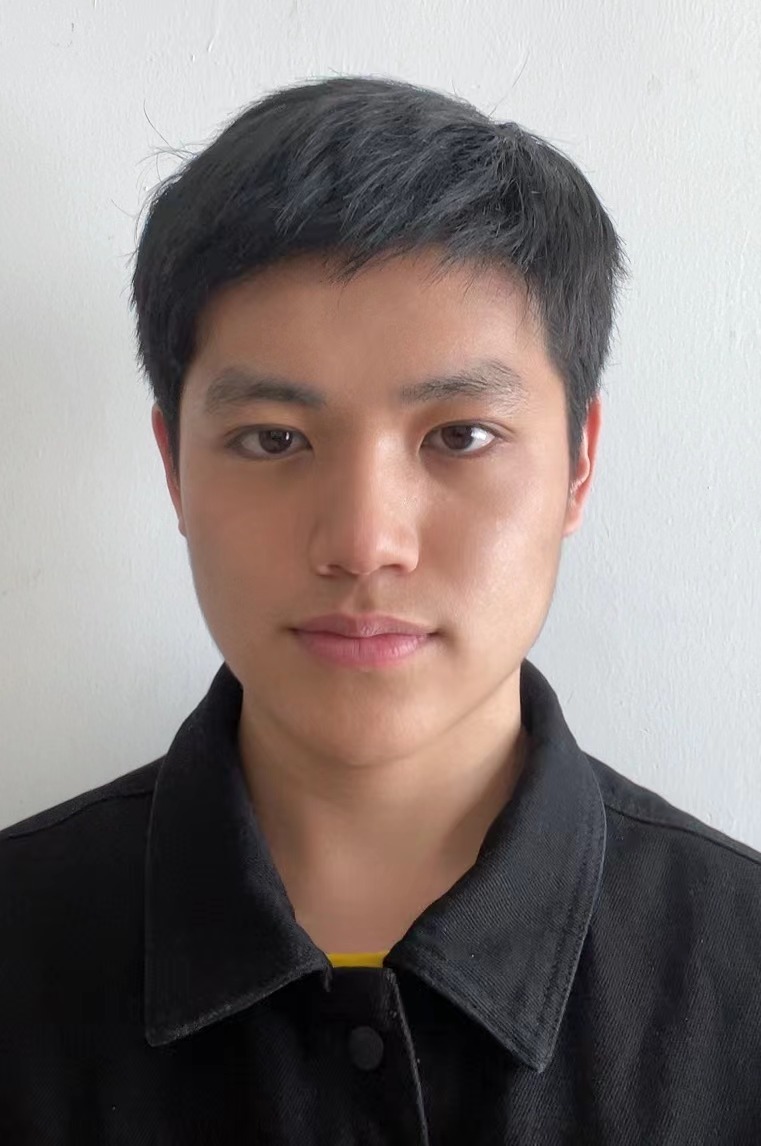}}]{Shuaiyu Xie}
 
 received his B.S. degree from the School of Computer Science, Wuhan University, China, in 2021. He is currently pursuing a Ph.D. degree at Wuhan University. His current research interests include services computing, artificial intelligence for IT operations (AIOps), and task scheduling.
\end{IEEEbiography}

\vspace{-5mm}

\begin{IEEEbiography}[{\includegraphics[width=1in,height=1.25in,clip,keepaspectratio]{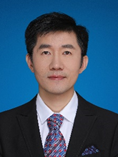}}]{Jian Wang}
 
received his Ph.D. degree in Computer Science from Wuhan University, China, in 2008. He is currently an Associate Professor in the School of Computer Science, Wuhan University, China. His current research interests include services computing and software engineering. He is now a member of the IEEE, a senior member of the China Computer Federation (CCF), and a member of the CCF Technical Committee on Services Computing (TCSC).
\end{IEEEbiography}

\vspace{-5mm}

\begin{IEEEbiography}[{\includegraphics[width=1in,height=1.25in,clip,keepaspectratio]{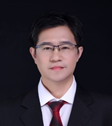}}]{Bing Li}
 
 received the Ph.D. degree from Huazhong University of Science and Technology, Wuhan, China, in 2003. He is currently a Professor at the School of Computer Science, Wuhan University. His main research areas are services computing, software engineering, artificial intelligence, and cloud computing.  He is now a member of the IEEE and a distinguished member of the China Computer Federation (CCF).
\end{IEEEbiography}

\vspace{-5mm}

\begin{IEEEbiography}[{\includegraphics[width=1in,height=1.25in,clip,keepaspectratio]{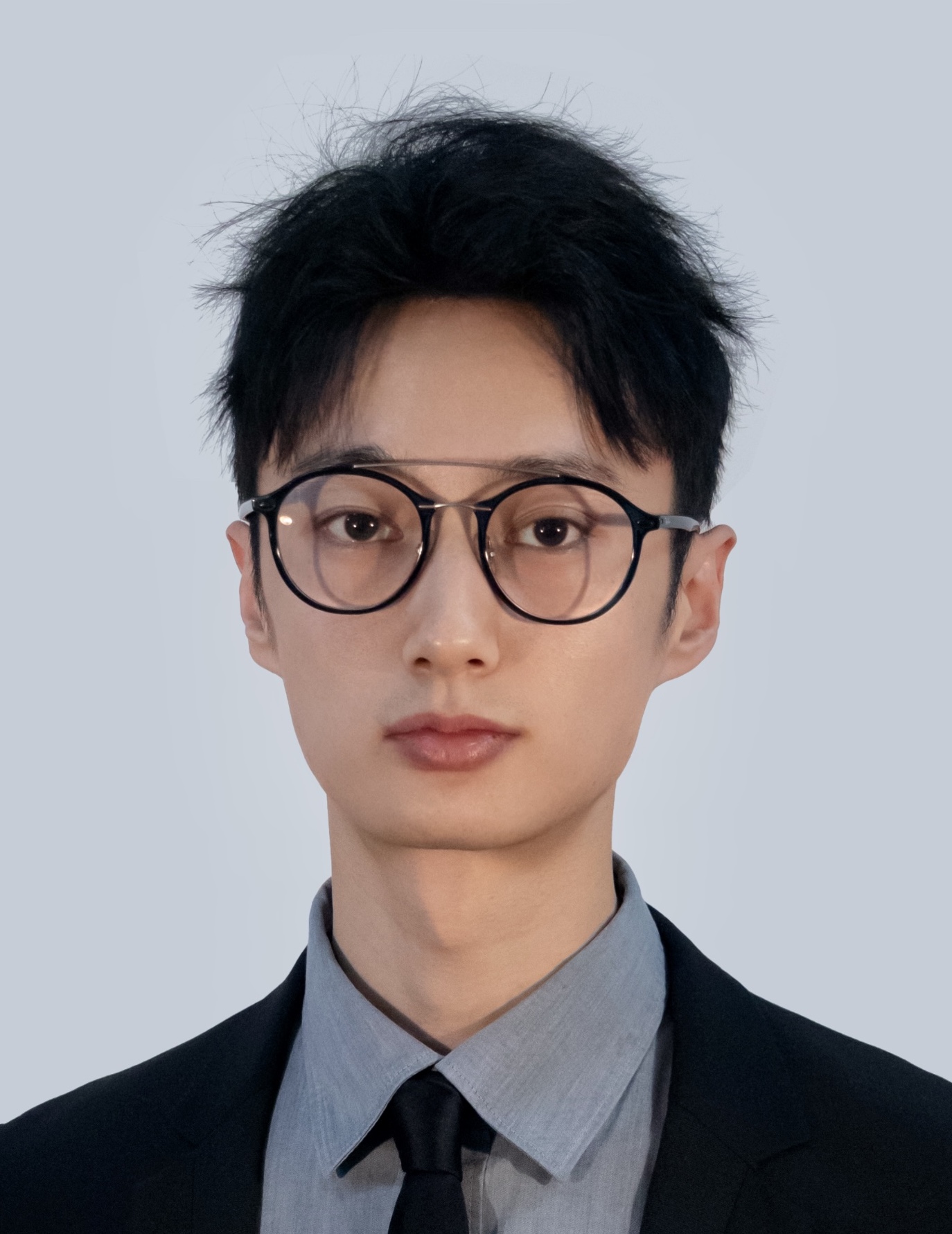}}]{Zekun Zhang}
 
received his M.S. degree from the school of computer science, Wuhan University, China. He is currently working toward the Ph.D. degree at the school of computer science, Wuhan University. His research interests include cloud computing and fault diagnosis.
\end{IEEEbiography}

\vspace{-5mm}

\begin{IEEEbiography}[{\includegraphics[width=1in,height=1.25in,clip,keepaspectratio]{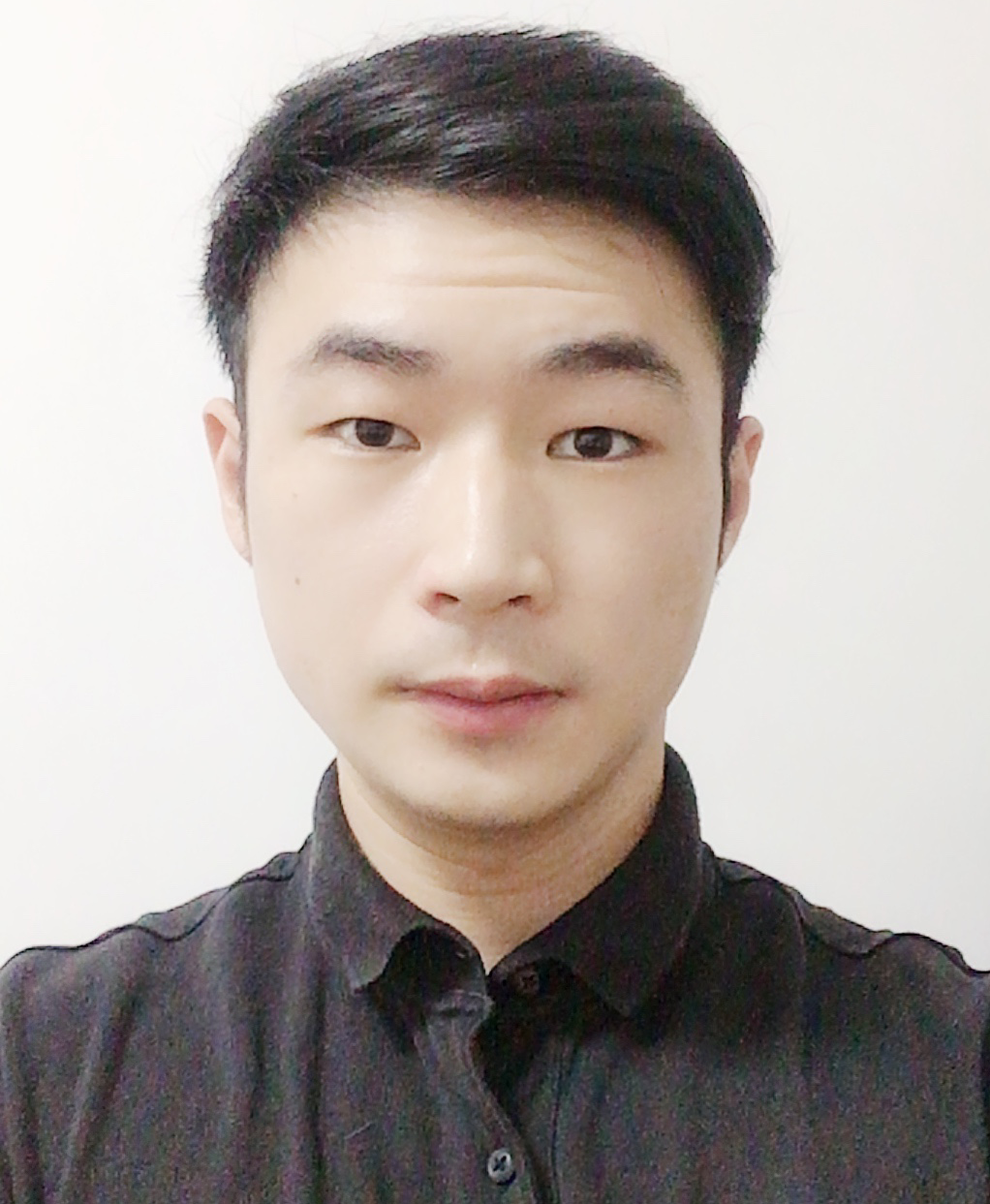}}]{Duantengchuan Li}
 
is currently working toward the Ph.D. degree in software engineering with the School of Computer Science, Wuhan University, Wuhan, China.
His research interests include recommendation system, representation learning, natural language processing, intention recognition, pattern recognition, computer vision and their applications in software engineering, intelligent education, and intelligent sports. 
\end{IEEEbiography}

\vspace{-5mm}

\begin{IEEEbiography}[{\includegraphics[width=1in,height=1.25in,clip,keepaspectratio]{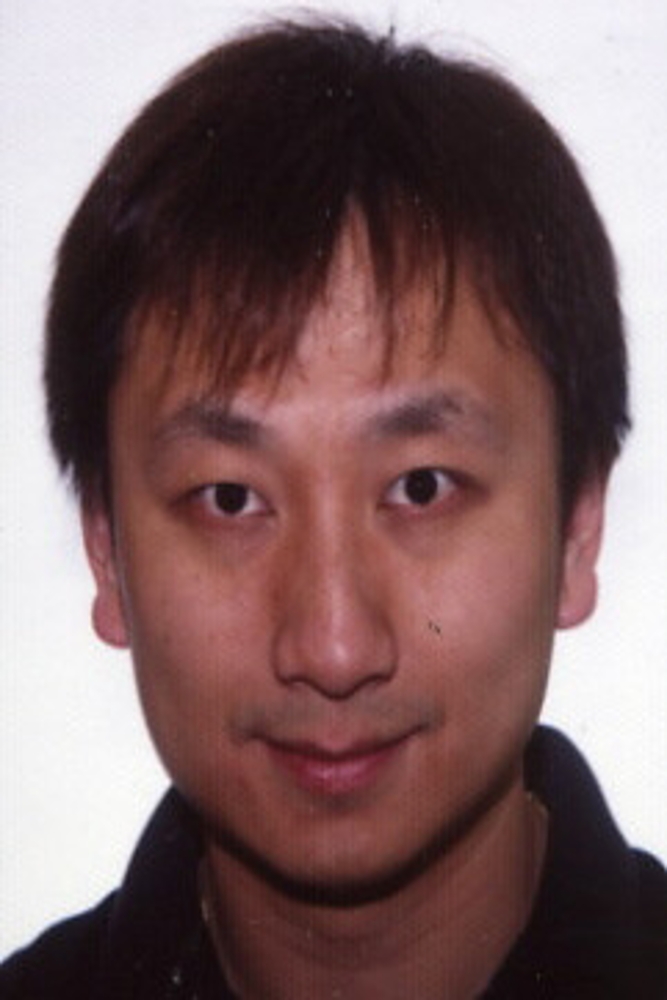}}]{Patrick C. K. Hung}
 
received the bachelor’s degree in computer science from the University of New South Wales, Sydney, NSW, Australia, in 1993, the master’s and Ph.D. degrees in computer science from Hong Kong University of Science and Technology, Hong Kong, in 1995 and 2001, respectively, and the master’s degree in management sciences from the University of Waterloo,  Canada, in 2002. He is a Professor and the Director of International Programs with the Faculty of Business and Information Technology, Ontario Tech University, Canada. His research interests include services computing, cloud computing, big data, and edge computing. He is a Founding Member of the IEEE Technical Committee on Services Computing and the IEEE Transactions on Services Computing.
\end{IEEEbiography}




\vfill

\end{document}